\documentstyle[epsf,psfrag]{elsart}
\renewcommand{\theequation}{\arabic{section}.\arabic{equation}}
\newcommand{\gtrsim}{ \raisebox{-0.5ex}{$\; \stackrel{>}{\sim} \;$}}
\newcommand{\lesssim}{\raisebox{-0.5ex}{$\; \stackrel{<}{\sim} \;$}}

\begin{document}
\begin{frontmatter}
%
%\draft
%
%
%            A B S T R A C T
%
%
\title{
Two-loop $\beta$-function from the exact renormalization group}
\author{Peter Kopietz}
\address{
Institut f\"{u}r Theoretische Physik der Universit\"{a}t G\"{o}ttingen,
Bunsenstrasse 9, 37073 G\"{o}ttingen, Germany\thanksref{Frankfurt}}
\thanks[Frankfurt]{Address after September 2000: Institut f\"{u}r Theoretische Physik,
Johann Wolfgang Goethe-Universit\"{a}t Frankfurt, Robert-Mayer-Strasse 10, 
60054 Frankfurt am Main.}
%
%\date{\today}
%
%\maketitle
%
\begin{abstract}
We calculate the  two-loop renormalization group (RG)
$\beta$-function of a massless scalar field theory 
from the irreducible version of
Polchinski's exact RG flow equation. 
To obtain the correct two-loop result
within this method, it is necessary to take
the full momentum-dependence of the irreducible
four-point vertex and the six-point vertex into account.
Although  the same calculation within the orthodox field theory method  
is less tedious, 
the flow equation method makes no assumptions
about the renormalizability of the theory,
and promises to be useful for performing two-loop calculations
for non-renormalizable condensed-matter systems.
We pay particular attention to the problem of the
field rescaling and the effect of the associated
exponent $\eta$ on the RG flow. 
\end{abstract}
\end{frontmatter}
%

%
%\narrowtext
%
%

\section{Introduction}
There are two different ways of formulating 
renormalization group (RG) transformations:
the first formulation \cite{ZinnJustin89}, 
which we shall call the field theory
method, relies on the fact that in a renormalizable field theory
all physical quantities can be expressed in terms of 
a finite number of renormalized couplings which
can be defined at an arbitrary scale $\mu$.
Because the bare theory is defined without reference to $\mu$, 
any change in the renormalized correlation
functions in response to a variation of $\mu$ must be compensated
by a corresponding change in the renormalized couplings, such that the
bare correlation functions remain independent of $\mu$.
This relation is usually expressed in terms of a partial 
differential equation for the renormalized correlation functions,
the Callan-Symanzik equation.
The Gell-Mann-Low  $\beta$-functions describe how the
renormalized couplings must depend on $\mu$ to guarantee that for 
fixed bare couplings the bare theory is independent of
$\mu$.
 
The other, more modern formulation of the RG was
pioneered by Wilson \cite{Wilson72}. In this approach, which has
been very fruitful in statistical physics and in condensed matter
theory \cite{Fisher98}, one starts from a local bare action
$S \{\phi \}$ with some ultraviolet cutoff $\Lambda_0$, and
integrates out the degrees of freedom for momenta down to some smaller
cutoff $\Lambda < \Lambda_0$.
After this elimination of the degrees of
freedom, the momenta and in general also the fields have 
to be properly rescaled
in order to obtain a fixed point of the RG \cite{Bell74}. 
In this way one obtains an effective
action with a new ultraviolet cutoff $\Lambda$.
In general, the new action
is a very complicated non-local functional of the new fields. 
The Wilsonian flow equations describe how the various couplings
of the effective action change as the scale
$\Lambda$ is reduced. 
An important  advantage 
of the Wilsonian RG over the field theory method is that
the Wilsonian method can also applied to non-renormalizable
field theories. In fact, condensed matter systems
are usually characterized by a finite lattice spacing,
which plays the role of a physical ultraviolet cutoff.
Hence, there is no need to perform the continuum limit
in condensed matter theory.

In the important work \cite{Wegner73}
Wegner and Houghton showed that  
the change of the effective action
due to an {\it{infinitesimal}} change in $\Lambda$ can be 
described in terms of a formally exact functional differential
equation. However, for practical calculations 
the Wegner-Houghton equation has some
unpleasant features, so that it has not been widely used to
solve problems of physical interest.  
Recently, there has been a renewed interest  
in exact formulations of the Wilsonian RG.
Polchinski \cite{Polchinski84} noticed that a particular
form of the exact Wilsonian RG equation (which is
now called the Polchinski equation, see Sec. \ref{subsec:Polchinski})
offers a straightforward
approach to proof the renormalizability of quantum field theories.

Due to its intuitive physical interpretation and its
greater generality, the Wilsonian RG is the method of
choice in condensed matter theory and statistical physics \cite{Fisher98}
as long as a one-loop calculation is sufficient.
In some problems, however,  
the correct behavior of physical observables
requires the knowledge of the two-loop RG $\beta$-function.
An example is the asymptotic 
low-temperature behavior of the susceptibility of 
classical \cite{Brezin76} and quantum \cite{Chakravarty89} 
Heisenberg magnets in two dimensions.
However, till now even condensed matter physicists resort
to the less intuitive  field theory method for
two-loop calculations \cite{Sachdev99}. In fact, 
the opinion seems to prevail that two-loop calculations within the
exact Wilsonian RG are more difficult than within the field theory
method. In this work we shall show that this is not necessarily
the case if one
uses the irreducible version of the exact RG \cite{Nicoll76,Chang92}
in the form derived by Wetterich \cite{Wetterich93} and, independently, 
by Morris \cite{Morris94}.

We consider a simple scalar field theory with a bare action
 \begin{equation}
 S \{ \phi \} = S_0 \{ \phi \} + S_{\rm int} \{ \phi \}
 \; ,
 \end{equation}
where the free part is 
 \begin{equation}
 S_0 \{ \phi \} 
%& = & 
%\frac{1}{2} \int d {\bf{r}} \phi ( {\bf{r}})
% G_0^{-1} ( {\bf{r}} - {\bf{r}}^{\prime} ) \phi ( {\bf{r}}^{\prime})
% \nonumber
% \\
  =  \frac{1}{2} \int_{ | {\bf{k}} | < \Lambda_0} 
\frac{ d {\bf{k}} }{ ( 2 \pi)^D}
 \phi_{\bf{k}} G_0^{-1} ( {\bf{k}}) \phi_{- {\bf{k}}}
 \; .
 \end{equation}
Here
 $\phi_{\bf{k}} = \int d {\bf{r}} e^{- i {\bf{k}} \cdot
 {\bf{r}} } \phi ( {\bf{r}})$ 
are the Fourier components of a real scalar field $\phi ( {\bf{r}} )$
in $D$ dimensions, and
 \begin{equation}
 G_0^{-1} ( {\bf{k}}) = {\bf{k}}^2 + m_0^2
 \; .
 \label{eq:G0kdef}
 \end{equation}
The interaction part $S_{\rm int} \{ \phi \}$ 
may also contain counter terms that are quadratic in the fields.
Later we shall
assume that it is a local function of the fields 
that is invariant under $\phi \rightarrow - \phi$.
The one-loop RG $\beta$-function for this theory has been
derived some time ago from the Polchinski equation \cite{Polchinski84} by 
Hughes and Liu \cite{Hughes88}.
Several strategies for calculating the RG $\beta$-function
at two loops have been proposed.
A derivation of the 
two-loop $\beta$-function entirely within the framework of the
exact RG was given by Papenbrock and Wetterich \cite{Papenbrock95},
who used the exact evolution equation for the 
effective average potential and worked with 
a smooth cutoff in momentum space.
Because sharp cutoffs in momentum space generate long-range
interactions in real space, Wilson and Kogut \cite{Wilson72} 
recommended to use a smooth cutoff
for calculations beyond one loop. On the other hand,
the one-loop $\beta$-function is
most conveniently performed with the Wilsonian RG
using a sharp cutoff, so that it   
would be useful to have a formulation of the Wilsonian RG
where the same cutoff procedure can also be
employed for higher order calculations.
The precise relation between the orthodox field theory
approach and the exact Wilsonian RG was established by
Bonini {\it{et al.}} \cite{Bonini97}, 
who used this relation to obtain the 
$\beta$-function at the two-loop order.
Another strategy for calculating the
two-loop  $\beta$-function of
$\phi^4$-theory was adopted   Pernici and Raciti \cite{Pernici98},
who derived the Gell-Mann Low equation
from the exact RG, and then used this equation
to express the $\beta$-function in terms of
Wilsonian Green functions. 
In Ref. \cite{Morris94} an attempt has been made to calculate
certain contributions to the two-loop coefficient of the
$\beta$-function by truncating the 
functional differential equations of the exact RG within
a momentum scale expansion; unfortunately,
this truncation is uncontrolled (see also Ref. \cite{Morris96}) and the result
is not a good approximation to the known two-loop result
obtained within the field theory method \cite{ZinnJustin89}.
Very recently Morris and Tighe \cite{Morris99}
succeeded in calculating the two-loop $\beta$-function 
from the exact RG by resumming the momentum scale expansion
for sharp cutoff and the derivative expansion for smooth cutoff
to infinite orders.
%To the best of our knowledge, a two-loop calculation of the
%RG $\beta$-function within the framework of the
%exact Wilsonian RG using a sharp cutoff in momentum space
%and without any reference to the Callan-Symanzik equation 
%or resumming momentum scale 
%or derivative expansions to infinite orders
%has not been performed.

In this work we shall show how 
the two-loop $\beta$-function can be obtained
directly within the framework of the exact Wilsonian RG using a sharp cutoff
in momentum space, without resumming the momentum scale
expansion, and  without any reference to the
Callan-Symanzik equation. 
Our motivation is to demonstrate that
a two-loop calculation within the exact RG
is conceptually quite simple and involves only
straightforward albeit tedious algebra. 
Our aim is to convince the reader that the exact RG is the most
convenient method for performing
two-loop RG calculations in condensed matter
theory and statistical physics.

The rest of this paper is organized as follows:
We begin in Sec.\ref{sec:different} with a brief
summary of the different versions of the
exact Wilsonian RG. In Sec.\ref{sec:rescaling} we discuss
the proper rescaling of momenta and fields in the exact RG. 
In Sec.\ref{sec:exact} we explicitly write down the
exact flow equations for the free energy, the
irreducible two-point vertex, the irreducible 
four-point vertex, and the irreducible six-point vertex.
We also
elaborate on the relation between the anomalous dimension of the
field and the irreducible two-point vertex.
In Sec.\ref{sec:general} we make some general remarks concerning the
structure of the RG flow. In Sec.\ref{sec:oneloop} we briefly re-derive
the one-loop $\beta$-function. The two-loop calculation is then
presented in Sec.\ref{sec:twoloop}. Our conclusions can be found in
Sec.\ref{sec:conclusions}. Finally, in an Appendix we briefly
summarize
the definition of various types of generating functionals and
introduce some convenient notations.

\section{Exact flow equations}
\setcounter{equation}{0}
\label{sec:different}

To integrate out degrees of freedom with momenta in the shell
$ \Lambda \leq | {\bf{k}} | < \Lambda_0$, we introduce the
cutoff-dependent free propagator $G_0^{\Lambda , \Lambda_0}$,
with matrix elements in momentum space given by
 \begin{equation}
 [G_{0}^{\Lambda, \Lambda_0}]_{ {\bf{k}} , {\bf{k}}^{\prime} } =
 ( 2 \pi )^D \delta ( {\bf{k}} - {\bf{k}}^{\prime} ) 
 G_0^{\Lambda , \Lambda_0} ( {\bf{k}}) 
 \; ,
 \end{equation}
where
 \begin{equation}
 G_0^{\Lambda , \Lambda_0} ( {\bf{k}}) =
 \frac{ \theta_{\epsilon} ( | {\bf{k}}| ,
   \Lambda) - \theta_{\epsilon} ( | {\bf{k}}| , \Lambda_0)}{ 
 {\bf{k}}^2 + m_0^2}
 \label{eq:G0cutoff}
 \; .
 \end{equation}
Here $\theta_{\epsilon} (  k  , \Lambda) \rightarrow 1$
for $k \gg \Lambda $, and
 $\theta_{\epsilon} ( k , \Lambda) \rightarrow 0$
for $k \ll \Lambda $, such that  $\theta_{\epsilon} ( k , \Lambda)$ 
drops from unity to zero mainly  
in the interval $ | k  - \Lambda | \leq 2 \epsilon$.
We are using here the same notation as Morris \cite{Morris94}.
For explicit calculations, we shall later on focus
on the sharp cutoff case,
% \begin{equation}
 $\lim_{\epsilon \rightarrow 0} \theta_{\epsilon} ( k , \Lambda)
 = \theta ( k  - \Lambda )$,
% \label{eq:cutoffdef}
% \; ,
% \end{equation}
but at this point, we shall keep $\epsilon$ finite.
There are several types of exact flow equations, 
corresponding to different types of correlation functions. 
A brief summary of the various generating functionals  and
their representations in terms of functional integrals or functional
differential operators is given in the Appendix. 
For convenience, let us now briefly summarize three types
of exact flow equations.
Our discussion closely follows Refs. \cite{Bagnuls00,Ellwanger94},
but we shall be more careful to keep track of all
field-independent contributions to
obtain the flow equation for the free energy.
Apart from the three flow equations given below,
two more types of flow equations should be mentioned:
The Wegner-Houghton equation \cite{Wegner73}, which 
was derived for system with discrete momenta and sharp cutoff,
is closely
related to the Polchinski equation \cite{Bagnuls00}.
Diagrammatically these  equations
contain terms which are not one-particle irreducible.
If one works with a sharp cutoff in momentum space, this
leads to ambiguities in the thermodynamic limit, where the momenta
form a continuum.  
Nicoll, Chang and collaborators \cite{Nicoll76}
avoided these difficulties by considering the flow equation 
for the Legendre effective action, which generates 
{\it{irreducible}} correlation functions. 
The exact flow equation given in
Sec.\ref{subsec:Polchinskir} is equivalent to
the flow equation of Ref. \cite{Chang92}.

\subsection{Polchinski equation for ${\cal{G}}_{\rm ac}^{\Lambda , \Lambda_0}$}
\label{subsec:Polchinski}

The  generating functional of the amputated cutoff-connected
Green functions ${\cal{G}}_{\rm ac}^{\Lambda , \Lambda_0}$ can
be defined via (see Eq. (\ref{eq:Gacdiff}))
 \begin{equation}
 e^{ {\cal{G}}_{\rm ac}^{\Lambda , \Lambda_0} \{ \phi \}}
 =  e^{ \frac{1}{2} ( \frac{\delta}{\delta \phi} , G_0^{\Lambda , \Lambda_0}
   \frac{\delta}{\delta \phi }) }
  e^{ - S_{\rm int} \{ \phi \}}
\label{eq:Gacdiff2}
 \; .
 \end{equation}
Differentiating both sides with respect to $\Lambda$, we obtain the
Polchinski equation \cite{Polchinski84},
 \begin{equation}
 \partial_{\Lambda} {\cal{G}}_{\rm ac}^{\Lambda , \Lambda_0}
 =  \frac{1}{2} \left[
 \left( \frac{\delta}{\delta \phi} , \partial_{\Lambda} G_0^{\Lambda ,
 \Lambda_0} \frac{\delta}{\delta \phi} \right) 
 {\cal{G}}_{\rm ac}^{\Lambda , \Lambda_0}
 + \left(  \frac{\delta {\cal{G}}_{\rm ac}^{\Lambda , \Lambda_0}}{
 \delta \phi} , \partial_{\Lambda} G_0^{\Lambda ,
 \Lambda_0}  \frac{\delta {\cal{G}}_{\rm ac}^{\Lambda , \Lambda_0}}{
 \delta \phi} \right) \right]
 \; ,
 \label{eq:Polchinskiac}
 \end{equation}
where $\partial_{\Lambda}$ on the right-hand side acts only
on $G_0^{\Lambda , \Lambda_0}$.
From Eq. (\ref{eq:Gacdiff2}) we see that the initial condition at
$\Lambda = \Lambda_0$ is
 \begin{equation}
 {\cal{G}}_{\rm ac}^{\Lambda_0 , \Lambda_0} \{ \phi \} = 
 - S_{\rm int} \{ \phi \}
 \; .
 \label{eq:initial}
 \end{equation}
Performing the functional derivatives in
Eq. (\ref{eq:Polchinskiac}) in Fourier space, the Polchinski equation
can also be written as
\begin{equation}
 \partial_{\Lambda} {\cal{G}}_{\rm ac}^{\Lambda , \Lambda_0}
 =  \frac{1}{2} \int \frac{ d {\bf{k}}}{ ( 2 \pi)^D}
  [\partial_{\Lambda} G_0^{\Lambda ,
 \Lambda_0} ( {\bf{k}}) ] 
 \left[  \frac{ \delta^{2} {\cal{G}}_{\rm ac}^{\Lambda , \Lambda_0} }{
 \delta \phi_{\bf{k}} \delta \phi_{- {\bf{k}}}}
 + \frac{ \delta {\cal{G}}_{\rm ac}^{\Lambda , \Lambda_0 } }{ \delta
 \phi_{\bf{k}}} 
 \frac{ \delta {\cal{G}}_{\rm ac}^{\Lambda , \Lambda_0 } }{ \delta
 \phi_{- \bf{k}}} \right]
 \; .
 \label{eq:Polchinskik}
 \end{equation}
This equation has been used in Ref. \cite{Hughes88} to derive
the one-loop $\beta$-function for our simple scalar field theory.
Let us emphasize again that diagrammatically the right-hand side of
Eq. (\ref{eq:Polchinskik}) contains terms that are one-particle
reducible. This leads to technical
difficulties in calculations beyond the leading order \cite{Morris94}.
Although this problem can be avoided by expanding  
${\cal{G}}_{\rm ac}^{\Lambda , \Lambda_0} \{ \phi \}$
in terms of so-called Wick-ordered monomials \cite{Salmhofer98},
for our two-loop calculation it will be more convenient to
start from  the exact flow equation for
the generating functional of the irreducible
correlation functions, see Eq. (\ref{eq:Polchinskir}) below.

\subsection{Flow equation for ${\cal{G}}_{ \rm c}^{\Lambda , \Lambda_0}$}

Before discussing in some detail the (for our
purpose) most convenient version of the exact RG,
let us mention (for completeness) the flow equation for the
generating functional  ${\cal{G}}_{ \rm c}^{\Lambda , \Lambda_0} \{ J \}$
of the connected Green functions,
which seems to be even less convenient than Eq. (\ref{eq:Polchinskiac}).
From Eq. (\ref{eq:Gacdef}) we see that
 ${\cal{G}}_{ \rm c}^{\Lambda , \Lambda_0} \{ J \}$ can be obtained from
that of the corresponding generating functional of the
amputated connected Green functions via
 \begin{equation}
 {\cal{G}}_{\rm c}^{\Lambda , \Lambda_0} \{ J \}
  =   {\cal{G}}_{\rm ac}^{\Lambda , \Lambda_0} \{ G_0^{\Lambda ,
      \Lambda_0} J \} + \frac{1}{2} ( J, G_0^{\Lambda , \Lambda_0} J )
 \label{eq:Gcdef2}
 \; .
 \end{equation}
The exact flow equation is
 \begin{eqnarray}
 \partial_{\Lambda} {\cal{G}}_{ \rm c}^{\Lambda , \Lambda_0}
 & =  & - \frac{1}{2} \Bigg[
 \left( \frac{\delta}{\delta J} , \partial_{\Lambda} (G_0^{\Lambda ,
 \Lambda_0})^{-1} \frac{\delta}{\delta J} \right) 
 {\cal{G}}_{\rm c}^{\Lambda , \Lambda_0}
 \nonumber
 \\
 & & \hspace{5mm} 
 + \left(  \frac{\delta {\cal{G}}_{\rm c}^{\Lambda , \Lambda_0}}{
 \delta J} , \partial_{\Lambda} (G_0^{\Lambda ,
 \Lambda_0})^{-1}  \frac{\delta {\cal{G}}_{\rm c}^{\Lambda , \Lambda_0}}{
 \delta J} \right)  + {\rm Tr} \partial_{\Lambda }  \ln
 G_0^{\Lambda , \Lambda_0 } \Bigg]
 \; ,
 \label{eq:Polchinskic}
 \end{eqnarray}
with initial condition at $\Lambda = \Lambda_0 $
 \begin{equation}
 {\cal{G}}_{ \rm c}^{\Lambda_0 , \Lambda_0}
 \{  J \} = 0
 \; .
 \label{eq:initialc}
 \end{equation}
This is an ill defined initial problem, so that
Eq. (\ref{eq:Polchinskic})
is not very useful in practice.

\subsection{Flow equation for ${\cal{G}}_{\rm ir}^{\Lambda , \Lambda_0}$}
\label{subsec:Polchinskir}

The exact RG equation for the
the generating functional of the irreducible vertices turns out
to be the  most convenient starting point for our two-loop calculation.
From Eq. (\ref{eq:Gammairdef2}) we have
 \begin{equation}
 {\cal{G}}_{\rm ir}^{\Lambda , \Lambda_0} \{ \varphi \}  =
( \varphi , J) -  \frac{1}{2} ( \varphi, (G_0^{\Lambda , \Lambda_0})^{-1} \varphi )
- {\cal{G}}_{\rm c}^{\Lambda , \Lambda_0} \{ J \{ \varphi
  \} \}
 \, ,
 \label{eq:Gammairdef3}
 \end{equation}
where $J$ is defined as a functional of $\varphi$ via
 \begin{equation}
 \varphi ( {\bf{r}}) = \frac{ \delta {\cal{G}}_{\rm c}^{\Lambda , \Lambda_0}
   \{ J \} }{\delta J ( {\bf{r}})}
 \; .
 \end{equation}
The flow equation for
$ {\cal{G}}_{\rm ir}^{\Lambda , \Lambda_0} \{ \varphi \}$
is \cite{Wetterich93,Morris94}
 \begin{eqnarray}
 \partial_{\Lambda} {\cal{G}}_{\rm ir}^{\Lambda , \Lambda_0}
  & = & \frac{1}{2} {\rm Tr} \left\{
 \partial_{\Lambda} (G_0^{\Lambda , \Lambda_0})^{-1} 
 \left[ ( G_0^{\Lambda , \Lambda_0} )^{-1} + {\cal{V}}^{ \Lambda,
     \Lambda_0}  \right]^{-1}  - \partial_{\Lambda} 
 \ln (G_0^{\Lambda , \Lambda_0})^{-1} \right\}
 \; ,
 \nonumber
 \\
 & &
 \label{eq:Polchinskir}
 \end{eqnarray}
where the functional ${\cal{V}}^{\Lambda, \Lambda_0} \{ \varphi \}$ 
is the second
functional derivative of ${\cal{G}}_{\rm ir}^{\Lambda , \Lambda_0}
 \{ \varphi \}$,
 \begin{equation}
 {\cal{V}}_{ {\bf{k}} , {\bf{k}}^{\prime }}^{ 
 \Lambda, \Lambda_0} \{ \varphi \}  =
 \frac{ \delta^2  {\cal{G}}_{\rm ir}^{\Lambda , \Lambda_0 } \{ \varphi
   \}}{ \delta \varphi_{\bf{k}} \delta \varphi_{{\bf{k}}^{\prime}}} \; .
 \label{eq:Hdef}
 \end{equation}
Unlike the authors of Refs. \cite{Wetterich93,Morris94},
we have  kept track of the field-independent
contributions to the right-hand side of Eq. (\ref{eq:Polchinskir}),
which determines the flow
of the free energy \cite{Wegner73}.
Note that in the non-interacting limit, i.e.
for ${\cal{V}}^{\Lambda , \Lambda_0} = 0$,
the right-hand side of Eq. (\ref{eq:Polchinskir}) vanishes.
The initial condition at $\Lambda = \Lambda_0$  is  simply
 \begin{equation}
 {\cal{G}}_{\rm ir}^{\Lambda_0 , \Lambda_0} \{ \varphi \}
 = S_{\rm int} \{ \varphi \}
 \; .
 \label{eq:initialGir}
 \end{equation}
It is convenient to separate at this point 
from Eq. (\ref{eq:Hdef}) the
field-independent part, which can be identified with
the irreducible self-energy,
  \begin{equation}
 {\cal{V}}^{ 
 \Lambda, \Lambda_0} \{ \varphi \}   =
 \Sigma^{\Lambda , \Lambda_0}  + 
 {\cal{U}}^{\Lambda, \Lambda_0}
 \{ \varphi \}
 \; ,
 \label{eq:Udef}
 \end{equation}
where by definition ${\cal{U}}^{\Lambda , \Lambda_0} \{ \varphi = 0 \}
= 0$. Then Eq. (\ref{eq:Polchinskir}) can also be written as
 \begin{eqnarray}
 \partial_{\Lambda} {\cal{G}}_{\rm ir}^{\Lambda , \Lambda_0}
  & =  & - \frac{1}{2} {\rm Tr} \Bigg\{
 \partial_{\Lambda} (G_0^{\Lambda , \Lambda_0})^{-1} 
 \Bigg[ ( G^{\Lambda , \Lambda_0} )^{2}  {\cal{U}}^{ \Lambda,
 \Lambda_0}   
 \left( 1 + G^{\Lambda , \Lambda_0} {\cal{U}}^{\Lambda , \Lambda_0}  
 \right)^{-1}
% \right.
% \right.
\nonumber
 \\ & &   
% \left. \left. 
 \hspace{25mm}
 + ( G_0^{\Lambda , \Lambda_0})^2 \Sigma^{\Lambda , \Lambda_0} 
 \left( 1 + G_0^{\Lambda ,
   \Lambda_0} \Sigma^{\Lambda , \Lambda_0} \right)^{-1} 
 \Bigg] \Bigg\}
 \; ,
 \label{eq:Pol3}
 \end{eqnarray}
where
 \begin{equation}
 G^{\Lambda , \Lambda_0} = [ (G_0^{\Lambda , \Lambda_0})^{-1} +
 \Sigma^{\Lambda , \Lambda_0} ]^{-1}
 \end{equation}
is the exact two-point function. Note that in momentum space
  \begin{equation}
 {\cal{V}}_{ {\bf{k}} , {\bf{k}}^{\prime }}^{ 
 \Lambda, \Lambda_0} \{ \varphi \}   =
 ( 2 \pi)^D \delta ( {\bf{k}} + {\bf{k}}^{\prime} )
 \Sigma^{\Lambda , \Lambda_0} ( {\bf{k}}) + 
 {\cal{U}}^{\Lambda, \Lambda_0}_{ {\bf{k}} , {\bf{k}}^{\prime}}
 \{ \varphi \}
 \; ,
 \label{eq:Udefkk}
 \end{equation}
 \begin{equation}
 G^{\Lambda , \Lambda_0}_{ {\bf{k}} , {\bf{k}}^{\prime}} =
 ( 2 \pi)^D \delta ( {\bf{k}} + {\bf{k}}^{\prime} )
 \frac{\theta_{\epsilon} ( | {\bf{k}}| ,
   \Lambda) - \theta_{\epsilon} ( | {\bf{k}}| , \Lambda_0 )}{
  {\bf{k}}^2 + m_0^2 + \Sigma^{\Lambda , \Lambda_0} ( {\bf{k}})}
 \label{eq:Gkk}
 \; .
 \end{equation}
For our two-loop calculation, it will be useful to take the sharp
cutoff limit of Eq. (\ref{eq:Pol3}). In this limit 
$\theta_{\epsilon} ( k , \Lambda ) \rightarrow
\theta ( k - \Lambda)$. 
Performing the trace in momentum space, 
Eq. (\ref{eq:Pol3}) reduces to \cite{Morris94}
 \begin{eqnarray}
 \partial_{\Lambda} {\cal{G}}_{\rm ir}^{\Lambda , \Lambda_0} 
  & = & - \frac{1}{2}
 \int \frac{ d {\bf{k}}}{ ( 2 \pi )^D} 
 \frac{\delta ( | {\bf{k}} | - \Lambda )  }{ \Lambda^2 + m_0^2 + 
 \Sigma^{\Lambda , \Lambda_0} ( {\bf{k}})}
 \left[ 
 {\cal{U}}^{ \Lambda, \Lambda_0}   
 \left( 1 + G^{\Lambda , \Lambda_0} {\cal{U}}^{\Lambda , \Lambda_0}  
 \right)^{-1} \right]_{ {\bf{k}} , - {\bf{k}}}
 \nonumber
 \\
 &  &
  - \frac{V}{2} \int \frac{ d {\bf{k}}}{ ( 2 \pi )^D} 
 \delta ( | {\bf{k}} | - \Lambda )  
\ln \left[  \frac{ \Lambda^2 + m_0^2 + \Sigma^{\Lambda ,
       \Lambda_0} ( {\bf{k}})}{ \Lambda^2 + m_0^2 } \right]
 \label{eq:Polsharp}
 \; ,
 \end{eqnarray}
where $V$ is the volume of the system.
In deriving Eq. (\ref{eq:Polsharp}) we have ignored
possible ambiguities that might arise from functions
$\theta (0)$  contained in the
term ${\cal{U}}^{ \Lambda, \Lambda_0}   
 \left( 1 + G^{\Lambda , \Lambda_0} {\cal{U}}^{\Lambda , \Lambda_0}  
 \right)^{-1}$.
See Ref. \cite{Morris94} and the remark in Sec.\ref{subsec:fourpoint}
for a further discussion of this point. 

\section{Rescaling  momenta and fields}
\setcounter{equation}{0}
\label{sec:rescaling}
 
As it stands, Eq. (\ref{eq:Polsharp}) describes the change in the
Legendre effective action due to the elimination of degrees of freedom
with momenta in the interval $[\Lambda , \Lambda_0 ]$. 
A complete  Wilsonian RG
transformation includes also a rescaling of momenta and
fields. A convenient way of including these rescaling
transformations into the  above flow equations it to rewrite them in
terms of dimensionless variables by multiplying all quantities by
appropriate powers of $\Lambda$. By taking the $\Lambda$-derivative
while keeping the dimensionless variables constant, 
the rescalings are implicitly carried out \cite{Bagnuls00}. 
In our case, it is useful to introduce dimensionless variables as follows,
 \begin{equation}
 {\bf{k}} = \Lambda {\bf{q}}
 \; \; , \; \;  
 \Lambda = \Lambda_0 e^{-t}
 \; \; , \; \; 
 \varphi_{\bf{k}} = \Lambda^{ D^{\varphi} - D } \sqrt{Z_t}
 \tilde{\varphi}_{\bf{q}}
 \; .
 \label{eq:tildephidef}
 \end{equation}
Here
 $D^{\varphi} = \frac{1}{2} ( D-2)$
is the scaling dimension of the field $\varphi ( {\bf{r}})$ in real
space (so that $D^{\varphi} - D$ is the scaling dimension of
its Fourier transform $\varphi_{\bf{k}}$). The scale-dependent
dimensionless factor $Z_{t}$ is the wave-function renormalization
factor,
which is related to the anomalous dimension $\eta_t $ of the
field via
 \begin{equation}
 \eta_t = - \partial_{t} \ln Z_t
 \; .
 \label{eq:etadef}
 \end{equation}
At this point the introduction of $Z_{t}$, which cannot be derived from
dimensional analysis, seems to be a rather ad hoc
procedure. However, in general it is necessary to
introduce such a factor, because otherwise it may happen that the
RG transformation never reaches a fixed point \cite{Bell74,Wegner73,Bagnuls00}.
Given Eq. (\ref{eq:tildephidef}), it is (almost) natural to
define
 \begin{equation}
 {\cal{G}}_{\rm ir}^{\Lambda , \Lambda_0} \{ \varphi_{\bf{k}} \}
 - \frac{N_t}{2} \ln Z_t
= {\cal{G}}_t \{ \tilde{\varphi}_{\bf{q}} \} 
 \; ,
 \label{eq:tildeGdef}
 \end{equation}
 \begin{equation}
 {\cal{U}}^{\Lambda , \Lambda_0} \{ \varphi_{\bf{k}} \}
 = \Lambda^{ - 2 D^{\varphi}} Z_t^{-1} 
 {\cal{U}}_{t}\{ \tilde{\varphi}_{\bf{q}} \} 
 \; ,
 \label{eq:tildeUdef}
 \end{equation}
 \begin{equation} 
G^{\Lambda , \Lambda_0} = \Lambda^{2 D^{\varphi}} Z_t G_t
 \; ,
 \label{eq:GZtdef}
 \end{equation}
 \begin{equation}
 {\bf{k}}^2 + m_0^2 + \Sigma^{\Lambda , \Lambda_0} ( {\bf{k}})
 = \Lambda^2 Z_t  r_t ( | {\bf{q}} | )
 \; ,
 \label{eq:rdef}
 \end{equation}
where have used the fact that $\Sigma^{\Lambda , \Lambda_0} (
{\bf{k}})$ depends only on $| {\bf{k}}|$. 
Here $N_t$ is the number of Fourier components in a system with
ultraviolet cutoff  $\Lambda = \Lambda_0 e^{-t}$, 
 \begin{equation}
 N_t = V \int \frac{ d {\bf{k}}}{ ( 2 \pi
   )^D} \theta ( \Lambda_0 e^{-t} - | {\bf{k}} | ) 
 = \frac{K_D}{D} V \Lambda_0^D e^{- Dt} \equiv N_0 e^{- Dt}
 \; ,
 \label{eq:Ntdef}
 \end{equation} 
where
$K_D$ is the volume of the unit sphere in $D$ dimensions divided
by $(2 \pi)^D$, 
 \begin{equation}
 K_D = \frac{ 1}{ 2^{D-1} \pi^{D/2} \Gamma ( D/2)}
 \; .
 \label{eq:Kddef}
 \end{equation}
The constant $\frac{N_t}{2} \ln Z_t$ in Eq. (\ref{eq:tildeGdef})
modifies the scaling of the free energy.
This term is due to the fact that the anomalous field rescaling
leads also to a change in the integration measure, as explained
in detail by Wegner and Houghton \cite{Wegner73}.
In our case, field rescaling  
$\phi_{\bf{k}}  \rightarrow \Lambda^{D^{\varphi} - D} Z_t \phi_{\bf{k}}$
modifies the generating functional of the cutoff connected
Green functions as
 \begin{equation}
 e^{ {\cal{G}}_c^{\Lambda , \Lambda_0} } \rightarrow 
  Z_{t}^{\frac{N_t}{2}} e^{ {\cal{G}}_c^{\Lambda , \Lambda_0}} 
 \; ,
 \end{equation}
so that
  \begin{equation}
  {\cal{G}}_c^{\Lambda , \Lambda_0}  \rightarrow 
  {\cal{G}}_c^{\Lambda , \Lambda_0} + \frac{N_t}{2} \ln Z_t 
 \; ,
 \end{equation}
and hence, from Eq. (\ref{eq:Gammairdef2})
  \begin{equation}
  {\cal{G}}_{\rm ir}^{\Lambda , \Lambda_0}  \rightarrow 
  {\cal{G}}_{\rm ir}^{\Lambda , \Lambda_0} - \frac{N_t}{2} \ln Z_t 
 \; .
 \label{eq:Gammirrescale}
 \end{equation}
Substituting the above rescaled variables into Eq. (\ref{eq:Polsharp})
the left-hand-side becomes (after multiplication with a
factor of $\Lambda$)
 \begin{equation}
 \Lambda \partial_{\Lambda}
 {\cal{G}}_{\rm ir}^{\Lambda , \Lambda_0} \{ \varphi_{\bf{k}} \}
 =  - \partial_t {\cal{G}}_t \{ \tilde{\varphi}_{\bf{q}} \}
 + \hat{\cal{R}}_{s} 
  {\cal{G}}_t \{ \tilde{\varphi}_{\bf{q}} \}
 + \frac{N_t}{2} \left[ \eta_t + D \ln Z_t \right]
 \; ,
 \label{eq:LdL}
 \end{equation}
where $\hat{\cal{R}}_s$ is the functional differential
operator describing the rescaling of fields and momenta \cite{Wegner73,Bagnuls00}
 \begin{eqnarray}
  \hat{\cal{R}}_{s}  & = &
  - \int \frac{ d {\bf{q}}}{ ( 2 \pi)^D } 
 \tilde{\varphi}_{\bf{q}} 
  {\bf{q}} \cdot \nabla_{\bf{q}}
 \frac{ \delta }{ \delta \tilde{\varphi}_{\bf{q}}}
 - {D}^{\varphi}_t 
 \int \frac{ d {\bf{q}}}{ ( 2 \pi)^D } 
 \tilde{\varphi}_{\bf{q}} 
 \frac{ \delta }{ \delta \tilde{\varphi}_{\bf{q}}}
 \nonumber
 \\
 &  = &
 \int \frac{ d {\bf{q}}}{ ( 2 \pi)^D } 
 \left[ ( {\bf{q}} \cdot \nabla_{\bf{q}} \tilde{\varphi}_{\bf{q}})
 + ( D - D^{\varphi}_t ) \tilde{\varphi}_{\bf{q}}
 \right]
 \frac{ \delta }{ \delta \tilde{\varphi}_{\bf{q}}}
 \; .
 \label{eq:Rsdef}
 \end{eqnarray}
Here 
 \begin{equation}
 D^{\varphi}_t = D^{\varphi} + \frac{\eta_t}{2} = \frac{1}{2}
 \left( D -2 + \eta_t \right)
 \end{equation}
is the full dimension of the renormalized field $\tilde{\varphi}$
(in real space).
Performing the angular integration in Eq. (\ref{eq:Polsharp})
in terms of $D$-dimensional spherical coordinates, we finally obtain
the exact flow equation 
 \begin{eqnarray}
\partial_t {\cal{G}}_t & = & \hat{\cal{R}}_{s} {\cal{G}}_t
 + \frac{K_D}{2 r_t ( 1)} \left\langle
 \left[ {\cal{U}}_t ( 1 + G_t {\cal{U}}_t )^{-1} \right]_{
   \hat{\bf{q}} , - \hat{\bf{q}}} \right\rangle_{\hat{\bf{q}}} 
 +  \frac{N_t}{2} \left[ \eta_t +  D \ln
 \left( \frac{ r_t (1)}{  r_0 ( 1)} \right) \right]
 \; .
 \nonumber
 \\
 \label{eq:Polt}
 \end{eqnarray}
Here $\langle \ldots \rangle_{\hat{\bf{q}}}$ means averaging over all
directions of the $D$-dimensional unit vector ${\hat{\bf{q}}}$.
Note that $G_t$ is a diagonal matrix in
momentum space with matrix elements
 \begin{equation}
 [ G_t ]_{ {\bf{q}} , {\bf{q}}^{\prime} } =
 ( 2 \pi)^D \delta ( {\bf{q}} + {\bf{q}}^{\prime})
 G_t ( {\bf{q}} )
 \; ,
 \end{equation}
 \begin{equation}
 G_t ( {\bf{q}}) = 
 \frac{ \theta ( | {\bf{q}} | - 1) - \theta ( | {\bf{q}} | - e^t) }{r_t ( | {\bf{q}} | )}
 \; ,
 \label{eq:Gtdef}
 \end{equation}
where $r_t ( q)$ is defined in Eq. (\ref{eq:rdef}).
The initial values at $t=0$ are by construction
 \begin{equation}
 G_{t=0} ( {\bf{q}}) = 0 \; \; , \; \; 
 Z_{t=0} = 1 \; \; , \; \; 
 r_{t=0} ( q ) = q^2 + \left( \frac{m_0}{\Lambda_0} \right)^2
 \; .
 \end{equation}
Eq. (\ref{eq:Polt}) is the central result of this section.
At a fixed point of the RG ${\cal{G}}_t$ become stationary,
 \begin{equation}
 \partial_t {\cal{G}}_t \{  \tilde{\varphi}_{\bf{q}} \} = 0
 \; .
 \label{eq:fixed}
 \end{equation}
Of course, usually it is impossible to solve Eq. (\ref{eq:Polt}) exactly, so
that we have to resort to approximations.

\section{Exact flow equations for the irreducible vertices}
\setcounter{equation}{0}
\label{sec:exact}

To generate an expansion in the number of loops, 
let us we expand the functional ${\cal{G}}_t \{ \tilde{\varphi}_{\bf{q}}\}$
in powers of Fourier components $\tilde{\varphi}_{\bf{q}}$ of the field,
 \begin{eqnarray}
 {\cal{G}}_t \{ \tilde{\varphi}_{\bf{q}} \}
 & = & \Gamma^{(0)}_{t} + \sum_{n= 1}^{\infty} \frac{1}{n !} 
\int \frac{ d {\bf{q}}_1 }{ ( 2
   \pi)^D} \ldots \int \frac{ d {\bf{q}}_n}{ ( 2 \pi)^D}
 ( 2 \pi )^D \delta ( {\bf{q}}_1 + \ldots + {\bf{q}}_n )
 \nonumber
 \\
 &  &  
 \hspace{10mm}
 \times
 \Gamma^{(n)}_{t} ( {\bf{q}}_1 , \ldots , {\bf{q}}_n )
 \tilde{\varphi}_{ {\bf{q}}_1 } \ldots \tilde{\varphi}_{ {\bf{q}}_n }
 \; .
 \label{eq:Gtexp}
 \end{eqnarray}
The functional ${\cal{U}}_t$ on the right-hand side of
Eq. (\ref{eq:Polt}) can be written as
 \begin{eqnarray}
 [ {\cal{U}}_t \{ \tilde{\varphi}_{\bf{q}} \}]_{ {\bf{q}} , {\bf{q}}^{\prime}} & = & 
 \sum_{n = 3}^{\infty} 
 \frac{1}{n !} 
 \int \frac{ d {\bf{q}}_1 }{ ( 2
   \pi)^D} \ldots \int \frac{ d {\bf{q}}_n}{ ( 2 \pi)^D}
 ( 2 \pi )^D \delta ( {\bf{q}}_1 + \ldots + {\bf{q}}_n ) 
 \nonumber
 \\
 &  &  \times
 \Gamma^{(n)}_{t} ( {\bf{q}}_1 , \ldots , {\bf{q}}_n )
 \frac{ \delta^2}{ \delta \tilde{\varphi}_{\bf{q}} \delta 
\tilde{\varphi}_{ {\bf{q}}^{\prime} } }
 \left[
\tilde{\varphi}_{ {\bf{q}}_1 } \ldots \tilde{\varphi}_{ {\bf{q}}_n }
 \right]
 \; .
 \label{eq:Utexp}
 \end{eqnarray}
We now substitute Eqs.(\ref{eq:Gtexp}) and (\ref{eq:Utexp})
into Eq. (\ref{eq:Polt}) and expand
 \begin{equation}
 {\cal{U}}_t ( 1 + G_t {\cal{U}}_t )^{-1}
 = {\cal{U}}_t - {\cal{U}}_t G_t {\cal{U}}_t +
 {\cal{U}}_t G_t {\cal{U}}_t G_t {\cal{U}}_t - \ldots
 \; .
 \label{eq:geomexp}
 \end{equation}
Collecting all terms with the same powers of the fields on both
sides of Eq. (\ref{eq:Polt}), 
we obtain an infinite hierarchy of flow equations for the
irreducible $n$-point vertices $\Gamma^{(n)}_t
({\bf{q}}_1 , \ldots , {\bf{q}}_n)$.
If the bare action is  invariant under $\varphi \rightarrow - \varphi$,
only the vertices with even $n$ are finite. 
It turns out that for the two-loop calculation of the RG $\beta$-function
it is sufficient to truncate the hierarchy by setting
$\Gamma^{(n)}_t
({\bf{q}}_1 , \ldots , {\bf{q}}_n ) = 0$ for $n \geq 8$.
We now explicitly give the
exact flow equations for the irreducible vertices that are 
required for the two-loop calculation.

\subsection{Free energy}

The flow equation for $\Gamma_t^{(0)}$ describes the flow of the
interaction correction to the free energy.
From the field-independent part of our exact flow equation
(\ref{eq:Polt}) we find
 \begin{equation}
 \partial_t \Gamma_t^{(0)} =  
\frac{N_t}{2} \left[ \eta_t + D \ln
 \left( \frac{ r_t (1)}{  r_0 ( 1)} \right) \right]
 \; .
 \label{eq:Gamma0flow}
 \end{equation}
The initial condition is
 \begin{equation}
 \Gamma_{t=0}^{(0)} = 0
 \; .
 \end{equation}
Setting $\Gamma_t^{(0)} = N_t f_t$ and using 
$N_t = N_0 e^{- Dt}$ (see
Eq. (\ref{eq:Ntdef})),
we obtain the flow equation for the interaction correction to the 
free energy per Fourier component,
 \begin{equation}
 \partial_t f_t = D f_t  
 + \frac{\eta_t}{2}  + \frac{D}{2} \ln
 \left( \frac{ r_t (1)}{  r_0 ( 1)} \right)
 \; .
 \label{eq:Gammafflow}
 \end{equation}
Note that in the corresponding flow equation
for the potential $v_0$ 
given by Wegner and Houghton (see Eq. (3.9) of Ref. \cite{Wegner73})
the term $\eta_t$ is replaced by $\eta - 2$, where
$\eta$ is assumed to be independent of $t$. 
The extra $-2$ is due to the fact that the potential
$v_0$ introduced in Ref. \cite{Wegner73} represents the 
total free energy, while our
$\Gamma^{(0)}_t$ contains only the interaction correction
to the free energy. Hence for a non-interacting system
our $\Gamma_t^{(0)}$
does not flow and remains identically zero.
The solution of Eq. (\ref{eq:Gammafflow}) with
the correct initial condition is
 \begin{equation}
 f_t = \frac{1}{2} \int_0^{t} d t^{\prime}
 e^{ D (t -t^{\prime})} \left[ \eta_{t^{\prime}} + D \ln
 \left( \frac{ r_{t^{\prime}} (1)}{  r_0 ( 1)} \right) \right]
 \; .
 \label{eq:Gammazerosol}
 \end{equation}

\subsection{Irreducible two-point vertex and self-energy}

From Eq. (\ref{eq:Polt}) we obtain 
the following exact flow equation for the
dimensionless irreducible two-point vertex,
 \begin{eqnarray}
 \partial_t \Gamma_t^{(2)} ( {\bf{q}} , - {\bf{q}} ) & = &
 \left( 2 - \eta_t -  {\bf{q}} \cdot \nabla_{\bf{q}} \right)
 \Gamma_t^{(2)} ( {\bf{q}} , - {\bf{q}} )
 \nonumber
 \\
 &  + &  \frac{K_D}{2r_t ( 1 )} \left\langle
 \Gamma^{(4)}_t ( {\bf{q}} , - {\bf{q}} , \hat{\bf{q}}^{\prime} , 
 - \hat{\bf{q}}^{\prime} ) \right\rangle_{\hat{\bf{q}}^{\prime}}
 \; .
 \label{eq:Gamma2flow}
 \end{eqnarray}
The initial condition at $t = 0$ is determined by the
interaction part of the bare action $S_{\rm int} \{ \varphi \}$, see
Eq. (\ref{eq:initialGir}).
If $S_{\rm int} \{ \varphi \}$ does not have any term quadratic
in the fields, then the initial condition is
% \begin{equation}
$ \Gamma_{t=0}^{(2)} ( {\bf{q}} , - {\bf{q}}) = 0$.
% \; .
% \label{eq:Gammainitial}
% \end{equation}
However, to reach the Gaussian fixed point in $D \geq 4$ 
it is necessary to fine tune the bare
action such that the system flows along the critical surface.
As explained in detail in  Sec.\ref{sec:general}, 
in this case the initial value $\Gamma_{t=0}^{(2)} ( 0,0)$
becomes a function of the bare four-point vertex
(see Fig.\ref{fig:flowgmu} and Eq. (\ref{eq:muginitial})). 
Of course, such  quadratic term in the fields can also be absorbed
into a redefinition of the Gaussian part of the action, but we find
it more convenient to include all interaction-dependent terms in
$S_{\rm int} \{ \varphi \}$.

By definition, our dimensionless parameter
$r_t ( q)$ is related to the irreducible two-point vertex
$\Gamma_t^{(2)} ( {\bf{q}} , - {\bf{q}})$ via 
 \begin{equation}
 r_t ( q )  =  Z_t r_0 ( q ) + \Gamma_t^{(2)} ( {\bf{q}} , - {\bf{q}})
% \\
% & = &  r_0 ( q ) + ( Z_t -1 ) r_0 ( q) + \Gamma_t^{(2)} ( {\bf{q}} , -
% {\bf{q}})
% \nonumber 
% \\
 \; ,
 \label{eq:rGamma2}
 \end{equation}
%  & = &
%Z_t \left[ r_0  ( q ) + \sigma_t ( q ) \right]
% \; .
% \label{eq:ru}
%\end{eqnarray}
implying
 \begin{equation}
 [ \partial_t -  2 +  \eta_t + {\bf{q}} \cdot \nabla_{\bf{q}}  ]
 [ \Gamma_t^{(2)} ( {\bf{q}} , - {\bf{q}} ) - r_t ( q) ]
% &  = &
% \nonumber
% \\
% & & \hspace{-57mm} 
%= [ \partial_t -  2 +  \eta_t + {\bf{q}} \cdot \nabla_{\bf{q}} ) ]
% r_t ( q) + 
 = 2 Z_t \frac{ m^2_0 }{ \Lambda_0^2 }
 \; .
% \nonumber
% \\
 \label{eq:rGamma2relation}
 \end{equation}
Hence, up to a re-definition of
the four-point vertex by a momentum-independent 
constant proportional to $m_0^2 /
\Lambda_0^2 $, the flow equations for $r_t ( q)$ 
and $\Gamma^{(2)}_t ( {\bf{q}} , -
{\bf{q}})$ are identical. 
Note that the authors of Ref. \cite{Chang92}
prefer to work with $r_t ( q)$ (their flow equation for $r_t ( q)$
is thus equivalent to our Eq. (\ref{eq:Gamma2flow}), except that
these authors replace $\eta_t$ by its fixed point value).
In this work we find it more convenient to  parameterize the two-point function
in terms of its irreducible part $\Gamma^{(2)}_t ( {\bf{q}} , - {\bf{q}})$.

Physically it is clear that
the irreducible two-point function should uniquely determine
the anomalous dimension $\eta_t = - \partial_t \ln Z_t$
of the field.
Hence,  $\eta_t$ on the right-hand
side of Eq. (\ref{eq:Gamma2flow}) implicitly depends
on $\Gamma^{(2)}_t ( {\bf{q}} , - {\bf{q}})$. 
The precise relation
becomes more transparent when  we rewrite Eq. (\ref{eq:Gamma2flow}) 
in terms of the dimensionless irreducible self-energy $\sigma_t ( q)$
defined by
 \begin{equation}
 \sigma_t ( q )  \equiv \Lambda^{-2}  
 \Sigma^{\Lambda , \Lambda_0} (  \Lambda {\bf{q}}) 
 =
 Z_t^{-1} \Gamma_t^{(2)} ( {\bf{q}} , - {\bf{q}}) 
 \; ,
 \label{eq:sigmatdef}
 \end{equation}
where $\Lambda = \Lambda_0 e^{-t}$. 
Using $\partial_t Z_t = - Z_t \eta_t$, it is easy to show 
that Eq. (\ref{eq:Gamma2flow}) is equivalent with
 \begin{eqnarray}
 \partial_t \sigma_t ( q ) & = &
 \left( 2  - {\bf{q}} \cdot \nabla_{\bf{q}} \right)
 \sigma_t (q) 
 \nonumber
 \\
 & + & \frac{K_D}{2 Z_t^2 [ r_0 ( 1 ) + \sigma_t ( 1) ]} \left\langle
 \Gamma^{(4)}_t ( {\bf{q}} , - {\bf{q}} , \hat{\bf{q}}^{\prime} , 
 - \hat{\bf{q}}^{\prime} ) \right\rangle_{\hat{\bf{q}}^{\prime}}
 \; .
 \label{eq:sigmaflow}
 \end{eqnarray}
The wave-function renormalization $Z_t$ on the right-hand side
is now fixed by demanding that \cite{Jungnickel96}
 \begin{equation}
 q^2 + \sigma_t ( q) -  \sigma_t ( 0 ) = Z_t^{-1} q^2 + O ( q^3)
 \; ,
 \label{eq:Ztdef1}
 \end{equation}
which implies
 \begin{equation}
 Z_t^{-1} =  1 + \left. 
 \frac{\partial \sigma_t ( q)}{ \partial (q^2) } \right|_{q^2 = 0} 
 \; .
 \label{eq:Ztdef}
 \end{equation}
From the definition Eq. (\ref{eq:sigmatdef}) it is then easy to show
that $Z_t$ is related to the irreducible two-point vertex via
  \begin{equation}
  Z_t = 1 - \left. \frac{ \partial \Gamma^{(2)}_t ( {\bf{q}} , -
   {\bf{q}})}{ \partial ( q^2) } \right|_{q^2 = 0}   
  \; .
 \label{eq:ZtGamma2}
 \end{equation}
Using Eq. (\ref{eq:Ztdef}) to eliminate $Z_t$ from Eq. (\ref{eq:sigmaflow}), we obtain
the following non-linear partial differential
equation for  $\sigma_t ( q)$,
\begin{eqnarray}
 \partial_t \sigma_t ( q ) & = &
 \left( 2  - {\bf{q}} \cdot \nabla_{\bf{q}} \right)
 \sigma_t (q) 
 \nonumber
 \\
 &   &  \hspace{-8mm} + \frac{K_D}{2  [ r_0 ( 1 ) + \sigma_t ( 1) ] }
 \left[ 1 + \left. 
 \frac{\partial \sigma_t ( q)}{ \partial (q^2) } \right|_{q^2 = 0} 
 \right]^{2}  \left\langle
 \Gamma^{(4)}_t ( {\bf{q}} , - {\bf{q}} , \hat{\bf{q}}^{\prime} , 
 - \hat{\bf{q}}^{\prime} ) \right\rangle_{\hat{\bf{q}}^{\prime}}
 \; .
 \label{eq:sigmaflow2}
 \end{eqnarray}
Given a solution of Eq. (\ref{eq:sigmaflow2}) with
appropriate initial condition,
the anomalous dimension of the field can be obtained from
Eq. (\ref{eq:Ztdef}).
Note, however, that the condition (\ref{eq:fixed})
that the RG has a fixed point requires
 \begin{equation}
 \partial_t \Gamma^{(2)}_t ( {\bf{q}} ,- {\bf{q}}) = 0
 \label{eq:fixed2}
 \; ,
 \end{equation}
which according to Eq. (\ref{eq:sigmatdef}) means that
 \begin{equation}
 \partial_t \sigma_t ( q ) = \eta \sigma_t ( q)
 \; ,
 \label{eq:sigmaeigen}
 \end{equation}
where $\eta $ is the value of $\eta_t$ at the RG fixed point.
Thus, for a system with a finite anomalous dimension the
irreducible self-energy is not stationary at a fixed point of the RG.
This is the reason why Eq. (\ref{eq:Gamma2flow}) is more useful than 
Eq. (\ref{eq:sigmaflow}) if we are interested in
possible fixed points of the RG. 
Keeping in mind that $\partial_t \sigma_t ( q)$
on the left-hand-side of Eq. (\ref{eq:sigmaeigen})
can be replaced by the right-hand side 
of Eq. (\ref{eq:sigmaflow2}), we see that Eq. (\ref{eq:sigmaeigen})
can also be viewed as a non-linear eigenvalue 
equation \cite{Bagnuls00,Comellas98}. 
From this point of view it is not surprising
that a RG fixed point can only be reached for certain discrete values
of $\eta$ \cite{Bell74}.

\subsection{Irreducible four-point vertex}
\label{subsec:fourpoint}

The irreducible four-point vertex satisfies the following flow
equation,
 \begin{eqnarray}
 \partial_t \Gamma^{(4)}_t ( {\bf{q}}_1 , {\bf{q}}_2 , {\bf{q}}_3 ,
 {\bf{q}}_4 ) & = & \left[ 4 - D - 2 \eta_t - \sum_{i = 1}^{4} 
 {\bf{q}}_i \cdot \nabla_{ {\bf{q}}_i } \right]
 \Gamma^{(4)}_t ( {\bf{q}}_1 , {\bf{q}}_2 , {\bf{q}}_3 ,
 {\bf{q}}_4 )
 \nonumber 
 \\
 &  & \hspace{-42mm} + \frac{K_D}{2 r_t ( 1 )} 
 \left\langle \Gamma^{(6)}_t ( \hat{\bf{q}} , - \hat{\bf{q}} ,
{\bf{q}}_1 , {\bf{q}}_2 , {\bf{q}}_3 , {\bf{q}}_4 )
 \right\rangle_{ \hat{\bf{q}}}
 \nonumber
 \\
 &   & \hspace{-42mm} - \frac{K_D}{ r_t ( 1)} 
 \Big\langle 
 \Gamma^{(4)}_t ( \hat{\bf{q}} , - \hat{\bf{q}} - {\bf{q}}_1 -
 {\bf{q}}_2  , {\bf{q}}_1 , {\bf{q}}_2 )
 G_t ( \hat{\bf{q}} + {\bf{q}}_1 + {\bf{q}}_2 )
 \Big.
 \nonumber
 \\
 & & \hspace{-30mm} \Big.
 \times \Gamma^{(4)}_t (  \hat{\bf{q}} + {\bf{q}}_1 +
 {\bf{q}}_2  , - \hat{\bf{q}} , {\bf{q}}_3 , {\bf{q}}_4 )
 + ( {\bf{q}}_2 \leftrightarrow {\bf{q}}_3 )
 + ( {\bf{q}}_2\leftrightarrow {\bf{q}}_4 )
 \Big\rangle_{\hat{\bf{q}}}
 \; .
 \label{eq:Gamma4flow}
 \end{eqnarray}
As pointed out by Morris \cite{Morris94}, this equation is only valid
if none of the combinations of external momenta ${\bf{q}}_1 + {\bf{q}}_2$,
${\bf{q}}_1 + {\bf{q}}_3$, and ${\bf{q}}_1 + {\bf{q}}_4$
that enter the propagators $G_t ( \hat{\bf{q}} + {\bf{q}}_1 +
{\bf{q}}_i)$
vanishes. Otherwise these expressions contain the function
$\theta ( 0)$, which is ambiguous. However,
in physical problems where nothing drastic
happens at vanishing momenta
these special points in momentum space should not matter
in the calculation of any observable. In this case we may
simply ignore this problem, which we shall do from now on. 
On the other hand, if the field acquires a vacuum expectation value so that
its ${\bf{q}}=0$ Fourier component is finite, the sharp cutoff limit
has to be taken more carefully, or the vacuum expectation
value has to be treated separately \cite{Morris94}.

\subsection{Irreducible six-point vertex}

Eq. (\ref{eq:Polt}) implies the following flow equation for
the irreducible six-point vertex,
 \begin{eqnarray}
 \partial_t \Gamma^{(6)}_t ( {\bf{q}}_1 , \ldots , {\bf{q}}_6 ) & = 
 & \left[ 6 - 2 D - 3 \eta_t - \sum_{i = 1}^{6} 
 {\bf{q}}_i \cdot \nabla_{ {\bf{q}}_i } \right]
 \Gamma^{(6)}_t ( {\bf{q}}_1 , \ldots , {\bf{q}}_6 )
 \nonumber 
 \\
 &  & \hspace{-36mm} + \frac{K_D}{2 r_t ( 1 )} 
 \left\langle \Gamma^{(8)}_t ( \hat{\bf{q}} , - \hat{\bf{q}} ,
{\bf{q}}_1 , \ldots ,  {\bf{q}}_6 )
 \right\rangle_{ \hat{\bf{q}}}
 \nonumber
 \\
 &   & \hspace{-36mm} - \frac{K_D}{ r_t ( 1)}
 \sum_{ \{I_1 , I_2 \}}^{15 \; {\rm terms}} 
 \Big\langle 
 \Gamma^{(4)}_t ( \hat{\bf{q}} , - \hat{\bf{q}} - {\bf{Q}}_1 , I_1 )
 G_t ( \hat{\bf{q}} + {\bf{Q}}_1  )
 \Gamma^{(6)}_t (  \hat{\bf{q}} + {\bf{Q}}_1 , - \hat{\bf{q}} , I_2 ) 
 \Big\rangle_{\hat{\bf{q}}}
 \nonumber
 \\
 & & \hspace{-36mm} + \frac{K_D}{ r_t ( 1)}
 \sum_{ \{I_1 , I_2 \}, I_3 }^{45 \; {\rm terms}} 
 \Big\langle 
 \Gamma^{(4)}_t ( \hat{\bf{q}} , - \hat{\bf{q}} - {\bf{Q}}_1 , I_1 )
 G_t ( \hat{\bf{q}} + {\bf{Q}}_1  )
 \Gamma^{(4)}_t ( \hat{\bf{q}} + {\bf{Q}}_1 , - \hat{\bf{q}} +
 {\bf{Q}}_2 , I_3  )
 \nonumber
 \\
 & & \hspace{-6mm} \times
 G_t ( \hat{\bf{q}} - {\bf{Q}}_2  )
 \Gamma^{(4)}_t (  \hat{\bf{q}} - {\bf{Q}}_2 , - \hat{\bf{q}} , I_2 ) 
 \Big\rangle_{\hat{\bf{q}}}
 \; .
 \label{eq:Gamma6flow}
 \end{eqnarray}
We are using here the same notation as Morris \cite{Morris94}:
${\bf{Q}}_i = \sum_{{\bf{q}}_k \in I_i} {\bf{q}}_k$,
and $\sum_{ \{ I_1 , I_2 \} , I_3, \ldots I_m}$ means a sum over all
disjoint subsets, $I_i \cap I_j = \emptyset$ such that $\cup_{i
  =1}^{m} I_i = \{ {\bf{q}}_1 , \ldots {\bf{q}}_n \}$.
The symbol $\{ I_1 , I_2 \}$ means that this  pair  is counted only
once. 
The $15$ terms of the sum involving the combination
$\Gamma^{(4)} \Gamma^{(6)}$ correspond to the
$15 = \tiny{ \left(  \begin{array}{c} 6 \\ 4 \end{array} \right)}$
possibilities of choosing distinct subsets of four momenta
out of the six available momenta.
The $45$ terms in the last sum of Eq. (\ref{eq:Gamma6flow})
correspond to one half of the
$ 90 = \tiny{\left( \begin{array}{c} 6 \\ 4 \end{array} \right)
\left( \begin{array}{c} 4 \\ 2 \end{array}  \right)}$
possibilities of picking subsets of four momenta out of a set of six
momenta, and then picking again subsets of 
two out of these four momenta. 

\section{Relevant and irrelevant couplings}
\setcounter{equation}{0}
\label{sec:general}

Before embarking on the two-loop calculation of the RG
$\beta$-function, we should identify the 
relevant and irrelevant couplings.

\subsection{Relevant couplings}

In  $ D \geq 4$ the only fixed point of the RG flow is the Gaussian
fixed point, where 
all irreducible $n$-point vertices with $n > 2$ vanish.
Following the usual jargon, we call all couplings with a positive or
vanishing scaling dimension {\it{relevant couplings}}.
For $D > 4$ there are only two relevant couplings. The first is
the momentum-independent part of the irreducible two-point vertex,
 \begin{equation}
 \mu_t \equiv \Gamma^{(2)}_t ( 0 , 0) = Z_t \sigma_t ( 0) 
 \label{eq:mut}
 \; .
 \end{equation}
This coupling is strongly relevant
and has scaling
dimension $+2$. The second relevant coupling
is the coefficient
of the term of order $q^2$ in the expansion of
$\Gamma^{(2)}_t ( {\bf{q}} , - {\bf{q}})$ for small dimensionless
wave-vectors ${\bf{q}}$,
 \begin{equation}
  c_t \equiv  \left. \frac{ \partial \Gamma^{(2)}_t ( {\bf{q}} , -
   {\bf{q}})}{ \partial ( q^2) } \right|_{q^2 = 0} = 1 - Z_t
  \; ,
 \label{eq:ctdef}
 \end{equation}
see Eq. (\ref{eq:ZtGamma2}).
%Note that with this definition the flowing anomalous dimension is
% \begin{equation}
% \eta_t = - \partial_t \ln ( 1 - c_t) = \frac{ \partial c_t}{ 1 - c_t}
% \; .
% \label{eq:etacrelation}
% \end{equation}
The scaling dimension of $c_t$ vanishes.
Let us emphasize that in general $c_t$ does not
vanish at the Gaussian fixed point, but approaches
a finite value which depends on the bare interaction,
see Fig.\ref{fig:flowgc} below.
Instead of $\mu_t$ and $c_t$, we could also parameterize
the two relevant couplings in terms of  $\sigma_t ( 0)$ and $Z_t$, 
but we prefer to work with $\mu_t$ and $c_t$, because
these quantities are  directly related to the derivatives of 
our dimensionless two-point vertex $\Gamma^{(2)}_t ( {\bf{q}} , - {\bf{q}})$.
%which become stationary at possible
%fixed points of the RG.
In $D \leq  4$ 
the momentum-independent
part of the four-point vertex
 \begin{equation}
 g_t \equiv \Gamma^{(4)}_t ( 0, 0, 0, 0)
 \label{eq:gtdef}
 \end{equation}
is also relevant.
The scaling dimension of $g_t$  is positive in $D < 4$ and vanishes in
$D =4$.
All other couplings are irrelevant in $D \geq 4$,  
so that
the RG trajectory in the infinite-dimensional parameter space of
all couplings rapidly flows approaches the three-dimensional subspace
spanned by the relevant couplings $\{ \mu_t , c_t , g_t \}$.
As explained in detail by Polchinski \cite{Polchinski84},
for sufficiently large $t$ the irrelevant couplings can then be
expanded in powers of the relevant couplings.

\begin{figure}
%\vspace{1cm}
\begin{center}
\psfrag{g_t}{$K_4 {g}_t$}
\psfrag{mu_t}{$\mu_t$}
\epsfysize8cm 
\hspace{5mm}
\epsfbox{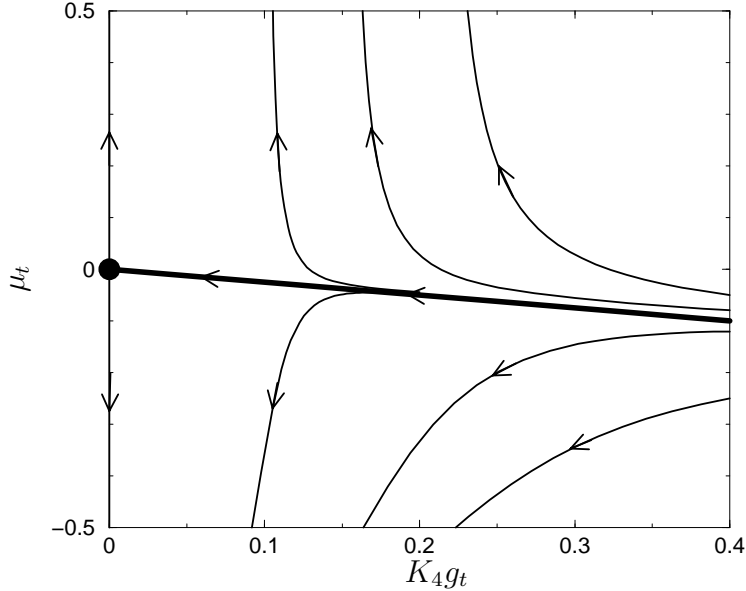}
\vspace{5mm}
\end{center}
\caption{
One-loop RG-flow for $D = 4$ in the $g$-$\mu$-plane.
Defining $\tilde{g}_t = K_D g_t$, the one-loop flow is determined by
$\partial_t \mu_t = 2 \mu_t + \frac{1}{2}  \tilde{g}_t/ ( 1 + \mu_t)$ and
$\partial_t \tilde{g}_t = (4 -D ) \tilde{g}_t - \frac{3}{2}  
\tilde{g}_t^2 /(1  + \mu_t)^2$. 
The thick black line is the linear approximation for the
critical trajectory ($\mu_t = - \frac{1}{4} \tilde{g}_t$, 
see Eq. (\ref{eq:mugrelation})), which flows towards  the Gaussian fixed
point (black dot). To reach this fixed point, the initial values 
$g_{t=0}$ and $\mu_{t=0}$ must be fine tuned to lie on the critical
trajectory.
}
\label{fig:flowgmu}
\end{figure}

From now on we shall consider  the flow on the critical surface
in $D \geq 4$, which is the infinite-dimensional manifold in coupling space
which flows into the Gaussian fixed point
under the RG transformation.
Recall that in  $D < 4$ the Gaussian
fixed point becomes unstable,
and a non-trivial fixed point emerges, 
the Wilson-Fisher fixed point \cite{Wilson72,Fisher98}.
For simplicity, we shall assume a vanishing bare mass ($m_0 =0$), so that
$r_0 ( q ) = {{q}}^2$.
The projection of the RG flow onto the
plane spanned by $g_t$ and $\mu_t$ is shown in Fig.\ref{fig:flowgmu}.
Obviously, at the Gaussian fixed point
the mass renormalization $\mu_t$
vanishes, so that  
this fixed point represents a system at criticality. 
For sufficiently small $g_t$ the critical trajectory
can be approximated by a straight line through the origin
in the $g$-$\mu$-plane,  
 \begin{equation}
 \mu_t = \gamma^{(1)} g_t + O ( g_t^2 )
 \; .
 \label{eq:mugrelation}
 \end{equation}
Note that Eq. (\ref{eq:mugrelation}) implies that, in order
to reach the Gaussian fixed point, the initial
value of the two-point vertex should be adjusted such that
 \begin{equation}
 \mu_{t = 0} \equiv \Gamma^{(2)}_{t=0} ( 0, 0 ) = \gamma^{(1)}
 \Gamma^{(4)}_{t=0} (0,0,0,0) 
 \; .
 \label{eq:muginitial}
 \end{equation}
To determine the numerical constant $\gamma^{(1)}$, we use the fact that
according to Eq. (\ref{eq:Gamma2flow})
 \begin{equation}
 \partial_t \mu_t = 2 \mu_t + \frac{K_D}{2 r_0 ( 1)} g_t
 + O ( g_t^2 )
 \; ,
 \label{eq:mug1}
 \end{equation}
while  Eq. (\ref{eq:Gamma4flow}) implies to leading order
 \begin{equation}
 \partial_t g_t = ( 4 - D ) g_t + O ( g_t^2)
 \; .
 \label{eq:gscale0}
 \end{equation}
Here we have anticipated that the lowest corrections
to  $\eta_t$ and $c_t = 1 - Z_t$ involve at least two powers
of $g_t$. Keeping in mind that
$r_0 ( 1 ) = 1$,
substituting Eq. (\ref{eq:mugrelation}) into
Eq. (\ref{eq:mug1}) and using Eq. (\ref{eq:gscale0}), 
we find
 \begin{equation}
 \gamma^{(1)} = - \frac{K_D}{ 2 ( D -2 ) }
 \; .
 \label{eq:gamma0res}
 \end{equation}
For the two-loop calculation of the $\beta$-function presented below
the linear accuracy given in Eq. (\ref{eq:mugrelation})
turns out to be sufficient.

Having adjusted the strongly relevant coupling
$\mu_t$ such that we reach the Gaussian fixed point, we are left with
two relevant couplings, $g_t$ and $c_t$. Of course, in addition
we have infinitely many irrelevant couplings, which are 
given by the higher order momentum-dependence
of the two-point and four-point vertices, and the higher 
irreducible vertices. However, the irrelevant couplings
can be expanded in powers of the relevant couplings, which 
is the key to perform a two-loop calculation within the flow
equation formalism.
Hence, in general we expect
that the projection of the exact RG flow 
onto the subspace spanned by couplings $g_t$ and $c_t$
can be described by equations of
the form
 \begin{equation}
 \partial_t c_t  =  A ( g_t , c_t ) 
 \; \; \; , \; \; \;
 \partial_t g_t   =  B ( g_t , c_t )
 \label{eq:gbeta}
 \; ,
 \end{equation}
with some dimensionless functions $A ( g, c)$ and $B(g,c)$.

\subsection{Irrelevant couplings} 

Because the RG trajectory 
rapidly approaches the manifold spanned by the 
relevant couplings, all couplings with negative scaling dimension
(the irrelevant couplings) become local functions
of the relevant couplings, which are independent of the initial
conditions at $t = 0$.  This is explained in detail
in the seminal paper by Polchinski \cite{Polchinski84}.
From the exact flow equation (\ref{eq:Polt}) for the irreducible vertices
it is easy to see that for small $g_t$ and large $t$
the leading behavior of the irrelevant coupling functions is
 \begin{equation}
 \Gamma_t^{(2)} ( {\bf{q}} , - {\bf{q}})
 - \mu_t - c_t q^2
  \sim   g_t^2 \gamma^{(2)} ( q )  + 
 O ( g_t^3 )
 \label{eq:Gammat2asym}
 \; .
 \end{equation}
  
 \begin{equation} 
  \Gamma_t^{(4)} ( {\bf{q}}_1 , {\bf{q}}_2 , {\bf{q}}_3 , {\bf{q}}_4 )
 - g_t
 \sim    g_t^2 
 \gamma^{(4)} ({\bf{q}}_1 , {\bf{q}}_2 , {\bf{q}}_3 , {\bf{q}}_4 ) 
 + O  ( g_t^3  )
 \label{eq:Gammat4asym}
 \; .
 \end{equation}

 \begin{equation}
  \Gamma_t^{(n)} ( {\bf{q}}_1 , \ldots , {\bf{q}}_n )
 \sim g_t^{n/2}
\gamma^{(n)}  ( {\bf{q}}_1 , \ldots , {\bf{q}}_n )  
 + O ( g_t^{ (n+2)/2} )
 \; .
 \label{eq:Gammat6asym}
 \end{equation}
Here the $\gamma^{(n)} ( {\bf{q}}_1 ,  \ldots , {\bf{q}}_n)$
are $t$-independent functions.
Note that the leading corrections are independent of
$c_t$. This is due to the fact that the leading term on the right-hand
side of the  flow equation (\ref{eq:gbeta}) for $c_t$ turns out
to be proportional to $g_t^2$, see Eq. (\ref{eq:Agcres}) below.
For a two-loop calculation within the exact RG
we need to know the functions $\gamma^{(4)} ( 
{\bf{q}}_1 , {\bf{q}}_2 , {\bf{q}}_3 , {\bf{q}}_4 )$ and
$\gamma^{(6)} ( {\bf{q}}_1 , \ldots , {\bf{q}}_6)$.

\section{One-loop $\beta$-function}
\setcounter{equation}{0}
\label{sec:oneloop}

The one-loop $\beta$-function has been calculated  
from Polchinski's flow equation by Hughes and Liu \cite{Hughes88}.
To this order, we may truncate the 
system of exact flow equation by setting
$\Gamma^{(n)}_t = 0$  for $n \geq 6$.
Furthermore, for a one-loop calculation
the momentum-dependence
of all couplings can be ignored.  In particular, 
the wave-function renormalization can be neglected at one-loop order,
so that   $Z_t = 1$ and hence $c_t = \eta_t = 0$. Obviously, the
function $A ( g_t , c_t )$
in Eq. (\ref{eq:gbeta}) vanishes to this order.
From the exact flow equation (\ref{eq:Gamma4flow})
we obtain for the momentum-independent part of the four-point vertex,
 \begin{equation}
 \partial_t g_t =  ( 4 - D) g_t - 3 K_D   
 \left\langle G_t ( \hat{\bf{q}} ) \right\rangle_{\hat{\bf{q}}}
 g_t^2
 + O ( g_t^3 , c_t g_t^2)
 \label{eq:oneloop1}
 \; ,
 \end{equation}
where $G_t ( {\bf{q}} )$ is defined in Eq. (\ref{eq:Gtdef}).
Note that by definition $G_{t = 0}  ( {\bf{q}}) = 0$, but for
$t > 0 $ and $| {\bf{q}} | \leq 1$ 
the ultraviolet cutoff $\theta ( | {\bf{q}} | - e^{t})$
in Eq. (\ref{eq:Gtdef})
can be neglected, so that to leading order we may approximate
 \begin{equation}
 G_t ( {\bf{q}} ) \approx  \frac{ \theta ( | {\bf{q}} | - 1 )}{ r_0 (
   | {\bf{q}} | )} \equiv C ( {\bf{q}})
 \; \; , \; \; t > 0 \; \; , \; \; | {\bf{q}} | \leq 1
 \; .
 \label{eq:Cqdef}
 \end{equation}
Using
 \begin{equation}
 \left\langle C ( \hat{\bf{q}} ) \right\rangle_{\hat{\bf{q}}} =
  \frac{1}{2}
 \; ,
 \label{eq:G0av}
 \end{equation}
we finally obtain 
for the  right-hand sides $A ( g_t , c_t)$ and $B ( g_t , c_t)$ 
of the flow equations (\ref{eq:gbeta}) for the relevant
couplings
 \begin{equation}
 A ( g_t , c_t) = O ( g_t^2 )
 \; ,
 \end{equation}
 \begin{equation}
 B ( g_t , c_t) = ( 4-D) g_t - \frac{3 }{2} K_D g_t^2
 + O ( g_t^3 )
 \; .
 \label{eq:Boneloop}
 \end{equation}
Eq. (\ref{eq:Boneloop}) is the known one-loop $\beta$-function of massless
$\phi^4$-theory \cite{ZinnJustin89}.
Recall that in four dimensions $K_4 = 2 / ( 4 \pi)^2$.

\section{Two-loop $\beta$-function}
\setcounter{equation}{0}
\label{sec:twoloop}

To obtain the flow equation for
$g_t$ at the two-loop order, we substitute
Eqs.(\ref{eq:Gammat4asym}) and (\ref{eq:Gammat6asym}) into
the exact flow equation
(\ref{eq:Gamma4flow}) for the irreducible four-point vertex
and collect all terms up to order $g_t^3$. After a straightforward
calculation we obtain for the generalized
$\beta$-function defined in Eq. (\ref{eq:gbeta}),
 \begin{eqnarray}
 B ( g_t , c_t ) & = & ( 4 - D  - 2 \eta_t ) g_t - \frac{3}{2} K_D g_t^2
\nonumber
 \\
 &   & \hspace{-17mm} + 3 K_D g_t^3   \left[ \gamma^{(1)}
 -   \left\langle \gamma^{(4)} ( \hat{\bf{q}} , -
   \hat{\bf{q}} , 0 , 0  ) \right\rangle_{\hat{\bf{q}}}
 + \frac{1}{6} \left\langle  \gamma^{(6)} ( \hat{\bf{q}} , -
   \hat{\bf{q}} , 0 , 0 , 0 , 0 ) \right\rangle_{\hat{\bf{q}}}
     \right] 
 \nonumber
 \\
 & & \hspace{-17mm}
+ O ( g_t^4)
 \label{eq:Btloop}
 \; .
 \end{eqnarray}
Recall that by definition
 \begin{equation}
 \eta_t = - \partial_t \ln ( 1 - c_t)  = \frac{ \partial_t c_t}{ 1 -
   c_t} 
 \; ,
 \label{eq:etatct}
 \end{equation}
so that to this order the function $B ( g_t , c_t)$
depends also on the second relevant coupling $c_t$.
The term $\gamma^{(1)}$ in the square braces of Eq. (\ref{eq:Btloop})
is due to the interaction-dependence of the mass renormalization
$\mu_t$ on the critical trajectory, see the thick solid arrow
in Fig.\ref{fig:flowgmu} and Eq. (\ref{eq:mugrelation}). 
The numerical value of $\gamma^{(1)}$ 
is given in Eq. (\ref{eq:gamma0res}).
The second term in the square brace of Eq. (\ref{eq:Btloop})
is due to the momentum-dependent part of the four-point vertex, while
the last term is due to the six-point vertex.
Hence, to calculate the RG flow at the two-loop order, 
we have to calculate the anomalous dimension to order $g_t^2$,
the momentum-dependent part of the four-point vertex to order $g_t^2$,
and the six-point vertex to order $g_t^3$. Note that it would be
incorrect to ignore the momentum-dependence of the six-point vertex,
because in Eq. (\ref{eq:Btloop}) we need 
$\Gamma^{(6)}_t ( {\bf{q}} , -
   {\bf{q}} , 0 , 0 , 0 , 0 )$ for $| {\bf{q}} | = 1$.

\subsection{Momentum-dependent part of four-point vertex}

Let us begin with the calculation of the momentum-dependent
part of the four-point vertex,
 \begin{equation}
 \tilde{\Gamma}^{(4)}_t ( {\bf{q}}_1 , {\bf{q}}_2 , {\bf{q}}_3 , {\bf{q}}_4)
 \equiv
 {\Gamma}^{(4)}_t ( {\bf{q}}_1 , {\bf{q}}_2 , {\bf{q}}_3 , {\bf{q}}_4)
 - g_t
  \label{eq:Gammatilde4}
 \; .
 \end{equation}
Assuming that the bare four-point vertex is  momentum-independent,
$\tilde{\Gamma}^{4}_t ({\bf{q}}_1 , {\bf{q}}_2 , {\bf{q}}_3 ,
{\bf{q}}_4 )$
satisfies the initial condition
at $t=0$,
 \begin{equation}  
 \tilde{\Gamma}^{(4)}_{t=0} ( {\bf{q}}_1 , {\bf{q}}_2 , {\bf{q}}_3 ,
 {\bf{q}}_4) = 0
 \; .
 \label{eq:initial4}
 \end{equation}
Subtracting from the exact flow equation (\ref{eq:Gamma4flow})
for the full four-point vertex the corresponding equation with all
external momenta set equal to zero, 
it is easy to show that to order $g_t^2$
the function
 $ \tilde{\Gamma}^{(4)}_t ( {\bf{q}}_1 , {\bf{q}}_2 , {\bf{q}}_3 ,
 {\bf{q}}_4)$ satisfies
 \begin{equation}
 \left[ \partial_t + D - 4 
 + \sum_{i=1}^4
 {\bf{q}}_i \cdot \nabla_{{\bf{q}}_i} \right]
 \tilde{\Gamma}^{(4)}_t ( {\bf{q}}_1 , {\bf{q}}_2 , {\bf{q}}_3 , {\bf{q}}_4)
 =   g_t^2 I^{(4)} ( {\bf{q}}_1 , {\bf{q}}_2 , {\bf{q}}_3 , {\bf{q}}_4  )
 \; ,
 \label{eq:flowGammatilde4}
 \end{equation}
where
 \begin{eqnarray}
 I^{(4)} ( {\bf{q}}_1 , {\bf{q}}_2 , {\bf{q}}_3 , {\bf{q}}_4  )
 & =  & - K_D \Big\langle C ( \hat{\bf{q}}^{\prime} + {\bf{q}}_1 + {\bf{q}}_2 )
 +  C ( \hat{\bf{q}}^{\prime} + {\bf{q}}_1 + {\bf{q}}_3 )
 \nonumber 
 \\
 & & 
  \hspace{7mm} +  C ( \hat{\bf{q}}^{\prime} + {\bf{q}}_1 + {\bf{q}}_4 ) - 3 C (
 \hat{\bf{q}}^{\prime}) \Big\rangle_{\hat{\bf{q}}^{\prime}}
 \label{eq:I4def}
 \; .
 \end{eqnarray}
On the right-hand side of Eq. (\ref{eq:I4def}) 
we have replaced the exact propagators
$G_t ( {\bf{q}} )$ by the non-interacting 
propagators without ultraviolet cutoff
$C ( {\bf{q}})$, see Eq. (\ref{eq:Cqdef}).
Such a replacement is only justified for $t \gtrsim 1$,
so that strictly speaking the approximation (\ref{eq:I4def})
cannot be used to study the flow of 
Eq. (\ref{eq:flowGammatilde4}) in the interval $0 \leq t \lesssim 1$.
However, as discussed by Polchinski \cite{Polchinski84},
the flow of irrelevant couplings for large $t$ becomes
independent of the precise initial conditions (see below), so that
for our purpose of investigating the 
infrared properties of the system we may ignore this subtlety. 
 
To solve Eq. (\ref{eq:flowGammatilde4}) with initial condition
(\ref{eq:initial4}), it is convenient to
introduce the auxiliary functions
 \begin{equation}
 \tilde{\Gamma}^{(4)}_{t,s} 
 ( {\bf{q}}_1 , {\bf{q}}_2 , {\bf{q}}_3 , {\bf{q}}_4)
 =
 {\Gamma}^{(4)}_t ( e^{-s} {\bf{q}}_1 , e^{-s } {\bf{q}}_2 , 
 e^{-s} {\bf{q}}_3 , e^{-s} {\bf{q}}_4)
 \; ,
 \label{eq:Gamma4s}
 \end{equation}
 \begin{equation}
 I^{(4)}_s (  {\bf{q}}_1 , {\bf{q}}_2 , {\bf{q}}_3 , {\bf{q}}_4  )
 = I^{(4)} (  e^{-s} {\bf{q}}_1 , e^{-s } {\bf{q}}_2 , 
 e^{-s} {\bf{q}}_3 , e^{-s} {\bf{q}}_4)
 \; .
 \label{eq:I4s}
 \end{equation}
Then the solution of Eq. (\ref{eq:flowGammatilde4}) 
is given by
   \begin{equation}  
 \tilde{\Gamma}^{(4)}_{t} ( {\bf{q}}_1 , {\bf{q}}_2 , {\bf{q}}_3 ,
 {\bf{q}}_4) = 
 \tilde{\Gamma}^{(4)}_{t, s = 0} ( {\bf{q}}_1 , {\bf{q}}_2 , {\bf{q}}_3 ,
 {\bf{q}}_4) 
 \; ,
 \label{eq:initial4tszero}
 \end{equation}
where
 \begin{equation}
 \left[ \partial_t  - \partial_s + D - 4  \right]
 \tilde{\Gamma}^{(4)}_{t,s} (  {\bf{q}}_1 , {\bf{q}}_2 , {\bf{q}}_3 , {\bf{q}}_4)
 =   g_t^2 I^{(4)}_s (  {\bf{q}}_1 , {\bf{q}}_2 , {\bf{q}}_3 , {\bf{q}}_4  )
 \; .
 \label{eq:tilde4s}
 \end{equation}
The initial condition (\ref{eq:initial4}) implies
  \begin{equation}  
 \tilde{\Gamma}^{(4)}_{t=0, s} ( {\bf{q}}_1 , {\bf{q}}_2 , {\bf{q}}_3 ,
 {\bf{q}}_4) = 0
 \; .
 \label{eq:initial4s}
 \end{equation}
Because in Eq. (\ref{eq:tilde4s}) the  external momenta
${\bf{q}}_i$ appear simply
as parameters, this equation
can be considered as a first order partial differential
equation with two variables, $t$ and $s$. The solution of 
Eq. (\ref{eq:tilde4s})
with the initial condition (\ref{eq:initial4s}) is easily obtained,
 \begin{equation}
 \tilde{\Gamma}^{(4)}_{t,s} 
 (  {\bf{q}}_1 , {\bf{q}}_2 , {\bf{q}}_3 , {\bf{q}}_4)
 = \int_0^{t} d \tau g^{2}_{\tau } e^{ - ( D - 4) (t - \tau )}
  I^{(4)}_{ s  + t - \tau } (  {\bf{q}}_1 , {\bf{q}}_2 , {\bf{q}}_3 , {\bf{q}}_4  )
 \; .
 \label{eq:tildeGamma4ssol}
 \end{equation}
Setting $s=0$ in Eq. (\ref{eq:tildeGamma4ssol}), we find
after a shift of the integration variable
  \begin{equation}
 \tilde{\Gamma}^{(4)}_t 
 (  {\bf{q}}_1 , {\bf{q}}_2 , {\bf{q}}_3 , {\bf{q}}_4)
 = \int_0^{t} d \tau g^2_{t - \tau } e^{ - ( D - 4) \tau }
  I^{(4)}_{\tau} (  {\bf{q}}_1 , {\bf{q}}_2 , {\bf{q}}_3 , {\bf{q}}_4  )
 \; .
 \label{eq:tildeGamma4res}
 \end{equation} 
At the first sight it seems that  
 $\tilde{\Gamma}^{(4)}_t 
 (  {\bf{q}}_1 , {\bf{q}}_2 , {\bf{q}}_3 , {\bf{q}}_4)$
is a non-local function of $g_t$. However,
by construction the 
term $ e^{ - ( D - 4) \tau }
  I^{(4)}_{\tau } (  {\bf{q}}_1 , {\bf{q}}_2 , {\bf{q}}_3 , {\bf{q}}_4  )$
in the integrand vanishes for large $\tau$ at least as fast as
 $e^{ - ( D - 3) \tau}$, so that the $\tau$- integration is
effectively cut off at $\tau \lesssim  1/(D-3)$.
For $t \gtrsim 1 / ( D-3)$ we may then 
approximate $g_{t - \tau } \approx g_t$ in the integrand
of Eq. (\ref{eq:tildeGamma4res}),
so that we finally arrive at the behavior anticipated
in Eq. (\ref{eq:Gammat4asym}),
 \begin{equation}
  \tilde{\Gamma}^{(4)}_t 
 (  {\bf{q}}_1 , {\bf{q}}_2 , {\bf{q}}_3 , {\bf{q}}_4)
% & \equiv & {\Gamma}^{(4)}_t 
% (  {\bf{q}}_1 , {\bf{q}}_2 , {\bf{q}}_3 , {\bf{q}}_4) - g_t
% \nonumber
% \\
  \sim   g_t^2 \gamma^{(4)} ( {\bf{q}}_1 , {\bf{q}}_2 , {\bf{q}}_3 ,
 {\bf{q}}_4 )
 \; \; , \; \; t \gtrsim 1  / ( D-3)
 \; ,
 \label{eq:tildeGamma4asym2}
 \end{equation}
with
 \begin{equation}
 \gamma^{(4)} ( {\bf{q}}_1 , {\bf{q}}_2 , {\bf{q}}_3 ,
 {\bf{q}}_4 ) = \int_0^{\infty } d \tau  e^{ - ( D - 4) \tau }
  I^{(4)}_{\tau} (  {\bf{q}}_1 , {\bf{q}}_2 , {\bf{q}}_3 , {\bf{q}}_4  )
 \; .
 \label{eq:gamma4res}
 \end{equation}
According to Eq. (\ref{eq:Btloop}),
for the two-loop  flow equation for $g_t$ we need 
 \begin{equation}
 \bar{\gamma}^{(4)}  \equiv 
 \left\langle \gamma^{(4)} ( \hat{\bf{q}} , -
   \hat{\bf{q}} , 0 , 0  ) \right\rangle_{\hat{\bf{q}}}
  =  \int_0^{1} d \lambda \lambda^{3 - D}
 \left\langle
 I^{(4)} ( \lambda \hat{\bf{q}} , - \lambda \hat{\bf{q}}, 0 , 0 )
 \right\rangle_{\hat{\bf{q}}} 
 \label{eq:bargamma4def}
 \; ,
 \end{equation}
where we have substituted $\lambda = e^{- \tau}$.
Note that the value of $\bar{\gamma}^{(4)}$
is determined by all momentum scales up to the infrared cutoff
$\Lambda$. The momentum scale expansion discussed by 
Morris \cite{Morris94,Morris96} corresponds to expanding the integrand
in Eq. (\ref{eq:bargamma4def}) to some finite order
in $\lambda$. Because the upper limit for the
$\lambda$-integration is $\lambda =1$, this expansion
is not controlled by a small parameter \cite{Morris94,Morris96}.
Obviously, the correct numerical value of the two-loop
coefficient of the $\beta$-function can only be obtained
if all terms in the momentum scale expansion are resummed \cite{Morris99}. 
Using the expression for $I^{(4)} ( {\bf{q}}_1 , {\bf{q}}_2 , 
{\bf{q}}_3 , {\bf{q}}_4 )$ given in Eq. (\ref{eq:I4def}) and
introducing $D$-dimensional spherical coordinates with
$\cos \vartheta = \hat{\bf{q}} \cdot \hat{\bf{q}}^{\prime}$,
we find
 \begin{eqnarray}
 \bar{\gamma}^{(4)} & = & - 2 K_D 
\frac{ \Omega_{D-1}}{\Omega_D} 
\int_0^{\pi} d \vartheta 
 ( \sin \vartheta )^{ D-2}
 \int_0^{1} d \lambda \lambda^{3- D}
 \nonumber
 \\
 & & \hspace{19mm }\times 
\left[
 \frac{ \theta ( \lambda + 2 \cos \vartheta) }{ 1 + 2 \lambda \cos
   \vartheta + \lambda^2 } - \theta ( \cos \vartheta ) \right]
 \; ,
 \label{eq:bargamma4int}
 \end{eqnarray}   
where $\Omega_D = ( 2 \pi )^D K_D$ is the volume of the unit sphere in
$D$ dimensions. Rearranging terms, 
$\bar{\gamma}^{(4)}$ can also be written as
 \begin{eqnarray}
 \bar{\gamma}^{(4)} & = &  2 K_D 
\frac{ \Omega_{D-1}}{\Omega_D}
 \left[ 
 \int_0^{\pi /2} d \vartheta 
 ( \sin \vartheta )^{ D-2}
 \int_0^{1} d \lambda 
 \frac{\lambda^{4-D} (
  2 \cos{\vartheta} + \lambda )}{1 + 2 \lambda \cos \vartheta + 
 \lambda^2}
 \nonumber
 \right.
 \\
 &  &
  \left. \hspace{18mm} -
\int_{0}^{\pi /6} d \vartheta 
 ( \cos \vartheta )^{ D-2}
 \int_{2 \sin \vartheta}^{1} d \lambda
 \frac{ \lambda^{3-D}}{1 - 2 \lambda \sin \vartheta + 
 \lambda^2}
 \right]
 \; ,
 \label{eq:bargamma4int2}
 \end{eqnarray}
where in the second integral we have shifted $\vartheta \rightarrow
\vartheta - \pi/2$. In Sec.\ref{subsec:four} we shall
explicitly evaluate this integral for $D=4$.
% \begin{eqnarray}
% \bar{\gamma}^{(4)} & = & 2 K_D 
% \frac{ \Omega_{D-1}}{\Omega_D}
% \left[
% \int_0^1 d \lambda
% \int_0^{1} d x \frac{( 1 - x^2)^{\frac{D-3}{2}} 
% (\lambda + 2 x) }{  \lambda^{D-4}  ( 1 + 2 \lambda x + \lambda^2 )}
% \nonumber \right.
% \\
% & & \left.
% - \int_{0}^{1} d\lambda  \int_{ - \lambda/2}^{0} d x
% \frac{( 1 - x^2)^{\frac{D-3}{2}}}{ \lambda^{ D-3} 
% (1 + 2 \lambda x + \lambda^2 )} \right]
% \; .
% \label{eq:bargamma4int2}
% \end{eqnarray}

\subsection{Anomalous dimension}

Having determined the momentum-dependent part of the four-point vertex
to order $g_t^2$, we can
calculate the momentum-dependent part of the two-point vertex to 
the same order. In analogy with Eq. (\ref{eq:Gammatilde4}), let us define
 \begin{equation}
 \tilde{\Gamma}^{(2)}_t ( q ) \equiv \Gamma^{(2)}_t ( {\bf{q}} , - {\bf{q}}) -
 \mu_t
 \; .
 \label{eq:Gammatilde2def}
 \end{equation}
Note that according to Eq. (\ref{eq:Gammat2asym})  we expect for large
$t$,
 \begin{equation}
 \tilde{\Gamma}^{(2)}_t ( q ) \sim c_t q^2 + g_t^2 \gamma^{(2)} ( q ) +
O ( g_t^3)
 \; .
 \label{eq:tildeGamma2asym}
 \end{equation}
Hence, the flow of our second
relevant coupling $c_t$ (which according to Eq. (\ref{eq:etatct})
determines the anomalous dimension $\eta_t$)
can be obtained from the quadratic term in the expansion of
$\tilde{\Gamma}^{(2)}_t ( q)$ for small $q$.
The function $\gamma^{(2)} ( q)$, which by construction
vanishes faster than $q^2$ for
small $q$, is not needed for a two-loop calculation.

Subtracting from the exact flow equation (\ref{eq:Gamma2flow})
for the irreducible two-point vertex
the same equation with ${\bf{q}}$ set equal to zero, we obtain to
leading order in $g_t$ 
 \begin{equation}
  \left[ \partial_t - 2 + {\bf{q}} \cdot \nabla_{\bf{q}} 
 \right] \tilde{\Gamma}^{(2)}_t ( q ) = g_t^2 I^{(2)} ( q ) + O ( g_t^3)
 \; ,
 \label{eq:tildeGammaflow}
 \end{equation}
where 
 \begin{equation}
 I^{(2)} ( q ) = \frac{K_D}{2} 
 \left\langle \gamma^{(4)} ( {\bf{q}} , - {\bf{q}} , \hat{\bf{q}}^{\prime}
   , - \hat{\bf{q}}^{\prime} ) -
 \gamma^{(4)} ( 0 , 0 , \hat{\bf{q}}^{\prime}
   , - \hat{\bf{q}}^{\prime} )  \right\rangle_{\hat{\bf{q}}^{\prime}}
 \label{eq:I2def}
 \; .
 \end{equation}
Expanding both sides of Eq. (\ref{eq:tildeGammaflow}) in powers of
$q$, we obtain to leading order
 \begin{equation}
 \partial_t c_t = \alpha_2 g_t^2 + O (g_t^3)
 \; ,
 \label{eq:flowctres}
 \end{equation}
where
 \begin{equation}
 \alpha_2 = \left. 
 \frac{ \partial I^{(2)} ( q )}{ \partial ( q^2)} \right|_{ q^2 = 0}
 \; .
 \label{eq:A2def}
 \end{equation}
From Eq. (\ref{eq:flowctres}) we conclude that the weak
coupling expansion of the function $A ( g_t , c_t)$ 
defined in Eq. (\ref{eq:gbeta}) is 
 \begin{equation}
 A ( g_t , c_t ) = \alpha_2 g_t^2 + O ( g_t^3, g_t c_t)
 \; .
 \label{eq:Agcres}
 \end{equation}
Eq. (\ref{eq:flowctres}) is easily integrated,
 \begin{equation}
 c_t = \alpha_2 \int_0^{t} d t^{\prime} g_{t^{\prime}}^2
 \; .
 \label{eq:ctprime}
 \end{equation}
Hence, the relation between $c_t$ and $g_t$ is non-local, so that
in general also
the anomalous dimension $\eta_t$ given in Eq. (\ref{eq:etatct})
is a non-local function of $g_t$. 
This is due to the fact that $c_t$ is not irrelevant.
However, to leading order
the term in the denominator of Eq. (\ref{eq:etatct}) can be neglected,
so that 
 \begin{equation}
 \eta_t = \partial_t c_t + O ( g_t^4) = \alpha_2 g_t^2 + O ( g_t^3)
 \; ,
 \label{eq:etares}
 \end{equation} 
i.e. to this order $\eta_t$ is a local function of $g_t$. 

For an explicit calculation of the 
number $\alpha_2$, we need to calculate the leading
$q$-dependence of the function $I^{(2)} (q)$.
Using Eqs.(\ref{eq:I2def}), (\ref{eq:gamma4res}) and   
(\ref{eq:I4def}), we find
 \begin{equation}
 I^{(2)} ( q ) = - K_D^2 \int_0^{1}  d \lambda \lambda^{3-D}
 \left\langle f ( \lambda | \hat{\bf{q}}^{\prime} + {\bf{q}} | )
 - f ( \lambda ) \right\rangle_{\hat{\bf{q}}^{\prime}}
 \; ,
 \label{eq:I2int}
 \end{equation}
where we have defined
 \begin{equation}
 f ( \lambda  )  =  \left\langle \frac{ \theta ( |
     \hat{\bf{q}}^{\prime \prime}  + \lambda \hat{\bf{e}} |) - 1}{ (
     \hat{\bf{q}}^{\prime \prime} + \lambda \hat{\bf{e}})^2 }
 \right\rangle_{ \hat{\bf{q}}^{\prime \prime}}
  =  \frac{\Omega_{D-1}}{\Omega_{D}} \int_{- \lambda/2}^{1} dx 
 \frac{( 1 - x^{2})^{\frac{D-3}{2}}  }{ 1 + 2 \lambda x + \lambda^2}
 \; .
 \label{eq:fdef}
 \end{equation}
Here $\hat{\bf{e}}$ is an arbitrary constant unit vector.
Apart from a different normalization, our function
$f ( \lambda)$ agrees with the corresponding function
introduced in Appendix D of Ref. \cite{Chang92}, were
the anomalous dimension at the Wilson-Fisher fixed point
has been calculated to order $( 4 - D)^2$.
Expanding $I^{(2)} ( q ) = \alpha_2 q^2 + O ( q^3)$ 
we find from Eq. (\ref{eq:I2int}) 
 \begin{equation}
 \alpha_2 = - \frac{K_D^2}{2 D} 
 \left[ (D-1) \int_0^{1} d \lambda \lambda^{4 - D}  f^{\prime} (
     \lambda) +
 \int_0^{1} d \lambda \lambda^{5 - D}  f^{\prime \prime} ( \lambda)
 \right]
 \; .
 \label{eq:A2res}
 \end{equation}
In Sec.\ref{subsec:four} we shall explicitly
evaluate Eq. (\ref{eq:A2res}) in four dimensions.

\subsection{six-point vertex}

To complete the two-loop calculation of the function
$B ( g_t , c_t)$ given in Eq. (\ref{eq:Btloop}), we 
need to know the number
 \begin{equation}  
  \bar{\gamma}^{(6)} =
 \left\langle  \gamma^{(6)} ( \hat{\bf{q}} , -
   \hat{\bf{q}} , 0 , 0 , 0 , 0 ) \right\rangle_{\hat{\bf{q}}}
 \; .
 \label{eq:bargamma6}
 \end{equation}
Recall that according to Eq. (\ref{eq:Gammat6asym})
the dimensionless function $\gamma^{(6)} ( {\bf{q}}_1 , \ldots ,
{\bf{q}}_6)$ is defined in terms of the large $t$-behavior
of the irreducible six-point vertex,
 \begin{equation}
  \Gamma_t^{(6)} ( {\bf{q}}_1 , \ldots , {\bf{q}}_n )
 \sim g_t^{3}
\gamma^{(6)}  ( {\bf{q}}_1 , \ldots , {\bf{q}}_n )  
 + O ( g_t^{4} )
 \; .
 \label{eq:Gammat6asym2}
 \end{equation}
From the exact flow-equation (\ref{eq:Gamma6flow})
we find that to order $g_t^3$ the irreducible six-point vertex
satisfies
 \begin{equation}
 \left[ \partial_t + 
 2 D - 6 + \sum_{i = 1}^{6} 
 {\bf{q}}_i \cdot \nabla_{ {\bf{q}}_i } \right]
 \Gamma^{(6)}_t ( {\bf{q}}_1 , \ldots , {\bf{q}}_6 ) = 
 g_t^3 I^{(6)} ( {\bf{q}}_1 , \ldots , {\bf{q}}_6 )
 \; ,
 \label{eq:Gamma6flow2}
 \end{equation}
where
 \begin{equation}
 I^{(6)} ( {\bf{q}}_1 , \ldots , {\bf{q}}_6 ) = K_D
 \sum_{ \{I_1 , I_2 \}, I_3 }^{45 \; {\rm terms}} 
 \Big\langle 
 C ( \hat{\bf{q}}^{\prime} + {\bf{Q}}_1  )
 C ( \hat{\bf{q}}^{\prime} - {\bf{Q}}_2  ) 
 \Big\rangle_{\hat{\bf{q}}^{\prime}}
 \; .
 \label{eq:I6def}
 \end{equation}  
Eq. (\ref{eq:Gamma6flow2}) 
can be solved with the same method
as Eq. (\ref{eq:flowGammatilde4}). The solution 
with initial condition
$ {\Gamma}^{(6)}_{t=0}
 (  {\bf{q}}_1 , \ldots  , {\bf{q}}_6) = 0$ is
  \begin{equation}
 {\Gamma}^{(6)}_t
 (  {\bf{q}}_1 , \ldots  , {\bf{q}}_6)
 = \int_0^{t} d \tau g^3_{t - \tau } e^{ - ( 2D - 6) \tau }
  I^{(6)} (   e^{ - \tau } {\bf{q}}_1 , \ldots , e^{- \tau} {\bf{q}}_6  )
 \; ,
 \label{eq:tildeGamma6res}
 \end{equation}
which is the analog of Eq. (\ref{eq:tildeGamma4res}).
For $t \gtrsim 1 /(2 D - 6)$ we may approximate
$g_{t - \tau} \approx g_t$ 
under the integral, so that
Eq. (\ref{eq:tildeGamma6res}) 
indeed reduces to Eq. (\ref{eq:Gammat6asym}), with
 \begin{equation}
 \gamma^{(6)} ({\bf{q}}_1 , \ldots  , {\bf{q}}_6 ) =  
 \int_0^{1} d \lambda \lambda^{2 D - 7} 
 I^{(6)} (   \lambda {\bf{q}}_1 , \ldots , \lambda {\bf{q}}_6  )
 \; ,
 \label{eq:gamma6res}
 \end{equation}
where we have substituted  $\lambda = e^{- \tau}$.
For our two-loop calculation we only need
$\gamma^{(6)} ( \hat{\bf{q}} , - \hat{\bf{q}}, 0,0,0,0)$.
Using Eq. (\ref{eq:I6def}) this can be written as
 \begin{eqnarray}
 \gamma^{(6)} ( \hat{\bf{q}} , - \hat{\bf{q}}, 0,0,0,0)
 & = & K_D \int_0^{1} d \lambda \lambda^{2 D - 7}
 \left[
  9 \left\langle C^2 ( \hat{\bf{q}}^{\prime })
  \right\rangle_{ \hat{\bf{q}}^{\prime}}
 \right.
 \nonumber
 \\
 &  & \hspace{-39mm} +
 \left. 12  \left\langle
 C ( \hat{\bf{q}}^{\prime} - \lambda {\hat{\bf{q}}}) 
 C ( \hat{\bf{q}}^{\prime})
 +
  C ( \hat{\bf{q}}^{\prime} + \lambda {\hat{\bf{q}}} ) 
 C ( \hat{\bf{q}}^{\prime})
 + 
 C^2 ( \hat{\bf{q}}^{\prime} + \lambda {\hat{\bf{q}}} ) 
 \right\rangle_{ \hat{\bf{q}}^{\prime}}
 \right]
 \label{eq:gamma6qq}
 \; .
 \end{eqnarray}
For the evaluation of $ \left\langle C^2 ( \hat{\bf{q}}^{\prime })
  \right\rangle_{ \hat{\bf{q}}^{\prime}}$
one should be careful to properly define the value of
the step function $\theta (x )$ at $x = 0$.
Following the procedure discussed by 
Morris \cite{Morris94}, we find
 \begin{equation}
 \left\langle C^2 ( \hat{\bf{q}}^{\prime })
  \right\rangle_{ \hat{\bf{q}}^{\prime}} = \frac{1}{3}
 \; .
 \label{eq:thetaorigin}
 \end{equation}
To obtain the number $\bar{\gamma}^{(6)}$ defined in
Eq. (\ref{eq:bargamma6}),
we should average Eq. (\ref{eq:gamma6qq}) over all directions of
the unit vector ${\hat{\bf{q}}}$.
Using $D$-dimensional spherical coordinates  we 
obtain
 \begin{eqnarray}
 \bar{\gamma}^{(6)} & = &   \frac{3 K_D }{ 2 ( D-3)}
 + 12 K_D \frac{ \Omega_{D-1}}{\Omega_D} 
 \int_{0}^{\pi} d \vartheta ( \sin \vartheta)^{D-2}
 \int_{0}^{1} d \lambda \lambda^{2 D-7}
 \nonumber
 \\
 & \times & 
 \Theta ( \lambda + 2 \cos \vartheta )  
 \left[ \frac{1}{1 + 2 \lambda \cos \vartheta + \lambda^2}
 + \frac{1}{ ( 1 + 2 \lambda \cos \vartheta + \lambda^2 )^2}
 \right]
% \right.
% \nonumber
% \\
% & + & 
% \left.
% \int_{0}^{\pi /6} d \vartheta ( \cos \vartheta)^{D-2}
% \int_{2 \sin \vartheta}^{1} d \lambda \lambda^{2 D-7}
%$ \left[ \frac{1}{1 - 2 \lambda \sin \vartheta + \lambda^2}
% + \frac{1}{ ( 1 - 2 \lambda \sin \vartheta + \lambda^2 )^2}
% \right]
% \right\}
 \; .
 \label{eq:bargamma6res}
 \end{eqnarray}
For an explicit evaluation of Eq. (\ref{eq:bargamma6res})
in $D=4$ see Sec.\ref{subsec:four}.

\subsection{$\beta$-function}

Collecting all contributions to the function $B ( g_t , c_t)$
given in Eq. (\ref{eq:Btloop}) and using the fact that
the dependence on $c_t$ enters only via
the anomalous dimension
 $\eta_t = \alpha_2 g_t^2 + O ( g_t^3)$, we see that at the two-loop order
the flow equation
(\ref{eq:gbeta}) for $g_t$ can be written as
 \begin{equation}
 \partial_t g_t = B ( g_t , c_t ) = \beta_0 g_t + \beta_1 g_t^2 + 
 \beta_2 g_t^3 + O ( g_t^4)
 \; ,
 \label{eq:betadef}
 \end{equation}
with
 \begin{equation}
 \beta_0  =   4  - D
 \; \; , \; \; 
 \beta_1  =  - \frac{3 K_D }{2}
 \; \; , \; \; 
 \beta_2  =  \beta_2^{\mu} + \beta_2^{\eta} + \beta_2^{(4)}
 + \beta_2^{(6)}
\label{eq:beta2res}
 \; ,
 \end{equation}
where
 \begin{eqnarray}
 \beta_2^{\mu} & = & 3 K_D \gamma^{(1)} = - \frac{3 K_D^2}{2 ( D-2)} 
 \label{eq:beta2mu}
 \; ,
 \\
 \beta_2^{\eta} & = & - 2 \alpha_2
 \label{eq:beta2etadef}
 \; ,
 \\
 \beta_2^{(4)} & = & -  3 K_D \bar{\gamma}^{(4)}
 \label{eq:beta24def}
 \; ,
 \\
 \beta_2^{(6)} & = & 
 \frac{K_D}{2} \bar{\gamma}^{(6)}
 \label{eq:beta26def}
 \; .  
 \end{eqnarray}
The term $\beta_2^{\mu}$ arises
from
flow of the momentum-independent part of the two-point vertex
along the critical surface, see Eq. (\ref{eq:gamma0res}).
The second term, $\beta_2^{\eta}$,
 is due to the flow of the anomalous dimension,
$\beta_2^{(4)}$ is due to the momentum-dependent part of the four-point vertex,
and $\beta_2^{(6)}$ is due to the six-point vertex.
Note that the numerical values of $\beta_2^{(4)}$ and $\beta_2^{(6)}$ 
are determined by the behavior of the four-point and
six-point vertices at finite momenta, and thus 
contain the effect of an infinite number of irrelevant couplings.

\subsection{Four dimensions}
\label{subsec:four}

We shall now explicitly evaluate the  two-loop coefficient
$\beta_2$ in $D=4$, and show that it agrees with the
result obtained within the field theory approach \cite{ZinnJustin89}.

According to Eq. (\ref{eq:beta2mu}),
the term $\beta_2^{\mu}$ is in $D=4$ given by
 \begin{equation}
 \beta_2^{\mu} = - \frac{3 K_4^2}{4} 
 \; \; , \;\; K_4 = \frac{2}{(4 \pi)^2} 
 \label{eq:beta2mures}
 \; .
 \end{equation}
Next, consider the contribution $\beta_2^{\eta}$
due to the anomalous dimension.
After an integration by parts we obtain
from Eq. (\ref{eq:A2res}) in $D=4$,
 \begin{equation}
 \alpha_2 = \frac{ K^2_4}{8}  \left[ 2 f (1) - 2 f (0)  + f^{\prime} ( 1)
 \right]
 \label{eq:A2D4}
 \; .
 \end{equation}
The required numbers are (see also Appendix D of Ref.\cite{Chang92}) 
 \begin{equation}
 f (0)  =  \frac{1}{2}
 \; \; , \; \;
 f ( 1)  =  \frac{2}{3} - \frac{\sqrt{3}}{ 2 \pi}
 \; \; , \; \; 
 f^{\prime} ( 1 )  =  - \frac{2}{3} + \frac{ \sqrt{3}}{\pi}
 \label{eq:prime1}
 \; ,
 \end{equation}
so that 
 \begin{equation}
 \alpha_2 = \frac{K_4^2}{24}  
 \; .
 \label{eq:A4res}
 \end{equation} 
Thus, the flowing anomalous dimension in $D = 4$ is
 \begin{equation}
 \eta_t =  \frac{K_4^2}{24}  g_t^2 + O ( g_t^3)
 \; ,
 \end{equation}
in agreement with the corresponding result
of the field theory method \cite{ZinnJustin89}.
From Eq. (\ref{eq:beta2etadef}) we conclude that
the contribution to the anomalous dimension
of the field to the two-loop coefficient
of the $\beta$-function  is
 \begin{equation}
 \beta_2^{\eta} = - \frac{K_4^2}{12}  
 \; .
 \label{eq:beta2eta}
 \end{equation}

The calculation of $\beta_2^{(4)} = - 3 K_D \bar{\gamma}^{(4)}$ 
is quite tedious. 
We first perform the $\lambda$-integrations in
the expression for $\bar{\gamma}^{(4)}$ given in
Eq. (\ref{eq:bargamma4int2}). In $D=4$ the 
integrals can be evaluated analytically,
 \begin{equation}
 \int_0^{1} d \lambda 
 \frac{
  2 \cos{\vartheta} + \lambda }{1 + 2 \lambda \cos \vartheta + 
 \lambda^2}
  = \frac{ \vartheta}{2}  \cot \vartheta + 
 \frac{1}{2} \ln [2(   1 + \cos \vartheta)] 
 \label{eq:lambdaint1}
 \; ,
 \end{equation}
 \begin{eqnarray}
 \int_{2 \sin \vartheta}^{1} d \lambda
 \frac{ 1 }{ \lambda (1 - 2 \lambda \sin \vartheta + 
 \lambda^2) } & =  & 
 \nonumber
 \\
 & & \hspace{-30mm}
\left( \frac{\pi}{4} - \frac{3 \vartheta}{2}
 \right) \tan \vartheta
 -  \frac{1}{2} \ln [8 ( 1 - \sin \vartheta ) ]  - \ln \sin \vartheta
 \label{eq:lambdaint2}
 \; .
 \end{eqnarray}
Substituting Eqs.(\ref{eq:lambdaint1}) and (\ref{eq:lambdaint2})
into Eq. (\ref{eq:bargamma4int2}), most of the
$\vartheta$-integrations can also be performed analytically.
After several integrations by parts the final result for
$\beta^{(4)}_2$ can be cast into the following form
 \begin{equation}
 \beta^{(4)}_2  =   K_4^2 \Bigg[
  - \frac{1}{2} + \frac{3 \sqrt{3}}{2 \pi} 
 - 2 \ln 2
  + \frac{6}{\pi} L \left( \frac{\pi}{3} \right) \Bigg] 
 \label{eq:beta4res}
 \; ,
 \end{equation}
%Here
% \begin{equation}
% C_{\rm atalan} = \sum_{k=0}^{\infty} \frac{(-1)^k}{ (2 k + 1)^2}
% = 0.915965 \ldots 
% \end{equation}
%is Catalan's constant \cite{Abramowitz72}, and
where 
 \begin{equation}
 L ( x ) = - \int_0^{x} d \vartheta \ln \cos \vartheta
 \label{eq:Ldef}
 \end{equation}
is known as Lobachevskiy's function \cite{Gradshteyn80}.
With $L ( \frac{\pi }{ 3}) =   0.218391 \ldots$
the numerical value of $\beta_2^{(4)}$ turns out to be
 \begin{equation}
 \beta_2^{(4)} \approx -  K_4^2 \times 0.642204 \ldots   
 \; .
 \end{equation}

\begin{figure}
\begin{center}
\psfrag{g}{$K_4 {g}_t$}
\psfrag{c}{$c_t$}
%\vspace{1cm}
\epsfysize8cm 
%\epsfysize5.3cm 
\hspace{5mm}
\epsfbox{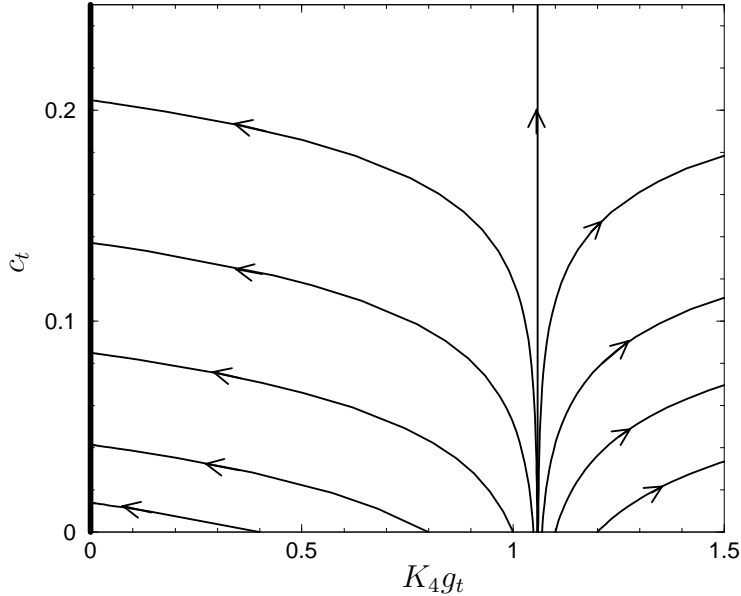}
\vspace{5mm}
\end{center}
\caption{
Two-loop RG-flow of the massless theory
for $D=4$ in the $g$-$c$ plane.
The flow is determined by $\partial_t c_t = \frac{1}{24}  \tilde{g}_t^2$ and
$\partial_t \tilde{g}_t = - \frac{3 }{2}  \tilde{g}_t^2 + \frac{17 }{12} 
\tilde{g}_t^3$, where $\tilde{g}_t = K_4 g_t$.
The thick line represents the Gaussian fixed point manifold, which can
be parameterized by the value of the 
wave-function renormalization $Z_{t = \infty} = 1 - c_{t = \infty}$,
or alternatively, the initial interaction $g_{t=0}$.
Note that in $D=4$ the two-loop $\beta$-function (\ref{eq:betadef})
vanishes at $K_4 {g}^{\ast} = \frac{18}{17} \approx
1.058823\ldots$.  
However, this is not a fixed point of the RG,
because for ${g}_{t} \geq {g}^{\ast}$ the second relevant 
coupling $c_t$ flows to strong coupling.
Because this graph is based on an expansion of the RG flow
equations in powers of $g_t$ and $c_t$,
only the weak coupling regime $ K_4 {g}_t \ll
1 $  and $c_t \ll 1$ can be trusted. 
}
\label{fig:flowgc}
\end{figure}

Finally, let us evaluate
the contribution $\beta_2^{(6)} = \frac{K_D}{2} \bar{\gamma}^{(6)}$
from the six-point vertex to the two-loop coefficient
of the $\beta$-function in $D=4$.
To calculate the integral
$\bar{\gamma}^{(6)}$ in Eq. (\ref{eq:bargamma6res}), it is
convenient to first integrate by parts. Writing
 \begin{equation}
 \frac{1}{ (1 + 2 \lambda \cos \vartheta + \lambda^2 )^2}
 = \frac{1}{ 2 \lambda \sin \vartheta} 
 \frac{\partial}{\partial \vartheta}
 \frac{1}{ 1 + 2 \lambda \cos \vartheta + \lambda^2 }
 \; ,
 \end{equation}
we obtain after some rearrangements
 \begin{eqnarray}
 \beta_2^{(6)} & = & K_4^2 \Bigg\{
 \frac{7}{4} + \frac{3 \sqrt{3}}{ 2 \pi}
  + \frac{6}{\pi} \int_0^{\pi /2} d \vartheta
 \int_0^{1} d \lambda \frac{ 2 \lambda \sin^2 \vartheta - \cos
   \vartheta}{ 1 + 2 \lambda \cos \vartheta + \lambda^2} 
 \nonumber
 \\
 & & \hspace{10mm} + 
 \frac{6}{\pi} \int_0^{\pi /6} d \vartheta
 \int_{2 \sin \vartheta}^{1} d \lambda \frac{ 2 \lambda \cos^2 
 \vartheta + \sin
   \vartheta}{ 1 - 2 \lambda \sin \vartheta + \lambda^2} 
 \Bigg\} 
\label{eq:beta6int2}
 \; .
 \end{eqnarray}
The $\lambda$-integrations can now be performed analytically,
 \begin{eqnarray}
 \int_0^{1} d \lambda \frac{ 2 \lambda \sin^2 \vartheta - \cos
   \vartheta}{ 1 + 2 \lambda \cos \vartheta + \lambda^2} 
 & = & - \frac{\vartheta}{2} \left[ \cot \vartheta + \sin (2 \vartheta ) \right]
 + \sin^2 \vartheta \ln  [ 2 ( 1 + \cos \vartheta)]
 \; ,
 \nonumber
 \\
 & &
 \label{eq:lambdaint3}
 \end{eqnarray}
 \begin{eqnarray}
 \int_{2 \sin \vartheta}^{1} d \lambda
 \frac{ 2 \lambda \cos^2 \vartheta + \sin \vartheta}{1 - 2 \lambda
   \sin \vartheta + \lambda^2 } & = &
 \nonumber
 \\
 & & \hspace{-30mm}
\left( \frac{\pi}{4} - \frac{3
     \vartheta}{2} \right) \left[ \tan \vartheta + \sin ( 2 \vartheta) \right]
 + \cos^2 \vartheta \ln [2 ( 1 - \sin \vartheta )]
 \label{eq:lambdaint4}
 \; .
 \end{eqnarray}
Substituting Eqs.(\ref{eq:lambdaint3}) and (\ref{eq:lambdaint4})
into Eq. (\ref{eq:beta6int2}) and performing the $\vartheta$-integrations
we obtain after some tedious re-shufflings
 \begin{equation}
 \beta^{(6)}_2  =   K_4^2 \Bigg[
   \frac{11}{4} - \frac{3 \sqrt{3}}{2 \pi} 
 + 2 \ln 2
  - \frac{6}{\pi} L\left( \frac{\pi}{3} \right) \Bigg] 
 \label{eq:beta6res}
 \; .
 \end{equation}
In deriving this expression,
we have used the fact that for 
$0 \leq x < \frac{\pi}{4}$
Lobachevskiy's function satisfies the functional 
relationship \cite{Gradshteyn80}
 \begin{equation}
 L ( x) - L \left( \frac{\pi}{2} - x \right)
 = \left( x - \frac{\pi}{4} \right) \ln 2
 - \frac{1}{2} L \left( \frac{\pi}{2} - 2 x \right)
 \; ,
 \label{eq:lobafunc}
 \end{equation}
which implies in particular
 \begin{equation}
 \frac{1}{2} \ln 2 
 - \frac{6}{\pi} L\left( \frac{\pi}{3} \right)
+ \frac{9}{\pi} L \left( \frac{\pi}{6} \right) = 0
 \; .
 \end{equation} 
We conclude that the contribution of the six-point vertex
to the two-loop coefficient of the $\beta$-function 
has the numerical value
 \begin{equation}
 \beta_2^{(6)} = K_4^2 \times 2.892204 \ldots
 \; .
 \label{eq:beta6num}
 \end{equation}
Neither $\beta_2^{(4)} / K_4^2$ nor $\beta_2^{(6)} / K_4^2$ is a 
rational number, but their sum is:
adding Eqs.(\ref{eq:beta4res}) and (\ref{eq:beta6res}), we find
 \begin{equation}
 \beta_2^{(4)} + \beta_2^{(6)} = \frac{9}{4} K_4^2 
 \label{eq:beta46add}
 \; .
 \end{equation} 
Collecting all terms, we obtain for the two-loop coefficient
of the RG $\beta$-function
 \begin{equation}
 \beta_2  = 
\beta_2^{\mu} + \beta_2^{\eta} + \beta_2^{(4)}
 + \beta_2^{(6)}
  =   \left[
 - \frac{3}{4} - \frac{1}{12} + \frac{9}{4} \right] K_4^2
 = \frac{17}{12} K_4^2
 \; ,
 \label{eq:finalbeta2}
 \end{equation}
which agrees with the result obtained by means of the field theory 
approach \cite{ZinnJustin89}. The two-loop RG-flow in the 
plane of the two relevant couplings $g_t$  and $c_t$ in $D=4$ is shown in
Fig.\ref{fig:flowgc}. 

\section{Conclusions}
\label{sec:conclusions}

In this work we have shown how the known two-loop results
for the RG $\beta$-function of a simple massless
scalar field theory can be obtained from the exact Wilsonian RG.
Although from a technical point of view the two-loop calculation
within the exact RG is
more tedious than the same calculation within the orthodox
field theory method \cite{ZinnJustin89}, the Wilsonian RG is conceptually
simpler, because there is no need
to invoke rather abstract concepts such as dimensional
regularization or minimal subtraction.
Moreover, the exact RG has the 
advantage that it can be used even if the problem of interest
cannot be  mapped onto a renormalizable field theory.
This is of particular importance in condensed matter systems,
where an underlying lattice spacing
always furnishes a physical ultraviolet cutoff, 
and there is no need for taking the continuum limit.
An interesting system where the orthodox field theory
approach fails is a two-dimensional electronic system
with a Fermi surface that contains saddle points \cite{Furukawa98}.
These give rise to van Hove singularities in the electronic density of
states, which in turn generate non-renormalizable singularities in 
perturbation theory. As a consequence, this problem cannot be
mapped onto a renormalizable field theory \cite{Gonzalez00}.

Due to its generality and conceptual simplicity,
the exact Wilsonian RG  
is very popular in condensed matter theory and statistical physics. 
However, even condensed matter theorists use the field theory method
for performing two-loop RG calculations \cite{Sachdev99}.
In the last few years the exact RG has been used to
study the fermionic many-body problem \cite{Shankar94}. 
To the best of our knowledge, 
two-loop RG calculations have not been performed for 
fermionic systems in $D > 1$.  But also for one-dimensional
fermionic systems, where the conventional field theory version of the
RG method has been applied for many years \cite{Solyom79}, 
an ambiguity inherent in the conventional way of 
performing two-loop calculations has recently been pointed 
out \cite{Carta00}. 
It seems to us that the exact Wilsonian RG, in its irreducible form 
used in this work, is the most promising method for performing
two-loop calculations for fermionic condensed matter systems.
This formulation of the exact RG
can be used even if the continuum limit does not exist,
and has the potential of clarifying
ambiguities of the type discussed in Ref. \cite{Carta00}.

\section*{Acknowledgments}

I am grateful to Boris Kastening, Tim Morris and Mario Raciti
for pointing out some relevant references.
I also thank Kurt Sch\"{o}nhammer for his comments on the manuscript.
This work was supported by the
Deutsche Forschungsgemeinschaft via a Heisenberg fellowship.

\appendix
\renewcommand{\theequation}{A.\arabic{equation}}
\renewcommand{\thesubsection}{A.\arabic{subsection}}

%\section*{Four types of generating functionals}
\section*{Appendix: Four types of generating functionals}

Here we briefly summarize some basic definitions and
representations of generating functionals for various
types of correlation functions. 
Although most of the
following equations can be found in 
textbooks \cite{ZinnJustin89}, we believe that this Appendix
greatly enhances the readability of this work by giving a concise
summary of the relevant expressions.

\subsection{Disconnected Green functions}

The generating functional of the disconnected Green functions is
given by the normalized partition function
in the presence of external sources, 
 \begin{equation}
 {\cal{G}} \{ J \} = \frac{ {\cal{Z}} \{ J \}}{ {\cal{Z}} \{0 \}}
 = \frac{\int {\cal{D}} \{\phi \} e^{ - S \{\phi\} + (J, \phi )}}{
 \int {\cal{D}} \{\phi \} e^{ - S \{\phi\} }}
 \label{eq:GJdef}
 \; ,
 \end{equation}
where we have used the notation
 \begin{equation}
 (J, \phi )  =  \int d{\bf{r}} J ( {\bf{r}}) \phi ( {\bf{r}})
  =  \int \frac{ d {\bf{k}}}{(2 \pi )^D} J_{{\bf{k}}} \phi_{- {\bf{k}}}
 \label{eq:scalardef}
 \; .
\end{equation}
The disconnected $n$-point functions are defined by
 \begin{equation}
 G^{(n)} ( {\bf{r}}_1, \ldots , {\bf{r}}_n) \equiv 
\langle \phi ( {\bf{r}}_1 ) \ldots \phi ( {\bf{r}}_n )
 \rangle
= \left. \frac{ \delta^{(n)}  {\cal{G}} \{ J \} }{ \delta J 
( {\bf{r}}_1)  \ldots \delta J ( {\bf{r}}_n)} \right|_{J = 0}
 \; .
 \end{equation}

\subsection{Connected Green functions}

The generating functional ${\cal{G}}_{\rm c} \{ J \}$
of the connected Green functions can be defined by
\begin{equation}
e^{{\cal{G}}_{\rm c} \{J \}} = 
 \frac{ {\cal{Z}} \{ J \}}{ {\cal{Z}}_0}
 =
 \frac{\int {\cal{D}} \{\phi \} e^{ - S \{\phi\} + (J, \phi )}}{
 \int {\cal{D}} \{\phi \} e^{ - S_0 \{\phi\} }}
 \; .
\label{eq:Gcdef}
\end{equation}
Note that by definition
$ - {\cal{G}}_{\rm c } \{J = 0 \} =  F - F_0$,
where $F$ is the free energy of the interacting system, and
$F_0 = - \ln {\cal{Z}}_0$ is the free energy of the non-interacting one.
% i.e.
% \begin{equation}
% e^{- F_0 } = {\cal{Z}}_0 = \int {\cal{D}} \{\phi \} e^{ - S_0 \{\phi\} }
% \; .
% \label{eq:Z0def}
% \end{equation}
The connected $n$-point functions can then be written as
   \begin{equation}
 G^{(n)}_{\rm c} ( {\bf{r}}_1, \ldots , {\bf{r}}_n) 
\equiv \langle \phi ( {\bf{r}}_1) \ldots \phi ( {\bf{r}}_n)
 \rangle_{\rm c }
= \left. \frac{ \delta^{(n)} {\cal{G}}_{\rm c} \{ J \} }{ \delta J ( {\bf{r}}_1)  \ldots \delta J (
  {\bf{r}}_n )} \right|_{J = 0}
 \label{eq:Gcndef}
 \; .
 \end{equation}
In contrast to Eq. (\ref{eq:GJdef}), we have normalized
the integral in Eq. (\ref{eq:Gcdef})
with the non-interacting partition function,
such that $- {\cal{G}}_{\rm c} \{ 0 \}$ is just the change of
the free energy due to the interaction. This normalization is useful,
because then ${\cal{G}}_{\rm c} \{ J \}$ can be written in terms of 
functional differential operators as follows,
 \begin{eqnarray}
 e^{ {\cal{G}}_{\rm c} \{ J \} }
 & = & e^{ -S_{\rm int} \{ \frac{\delta}{\delta J} \} } 
 e^{ \frac{1}{2} ( J, G_0 J )}
 \nonumber
 \\
 & = & e^{ \frac{1}{2} ( J, G_0 J )} 
 \left[ e^{ \frac{1}{2} ( \frac{\delta}{\delta \phi} , G_0
   \frac{\delta}{\delta \phi }) }
  e^{ - S_{\rm int} \{ \phi \}} \right]_{ \phi = G_0 J}
 \; .
\label{eq:Gcrepres}
 \end{eqnarray}

\subsection{Amputated connected Green functions}

The generating functional of the amputated
connected Green functions is
 \begin{eqnarray}
 {\cal{G}}_{\rm ac} \{ {\phi} \}
 & = & \left[ {\cal{G}}_{\rm c } \{ J \} - \frac{1}{2} ( J, G_0 J ) \right]_{ J =
   G_0^{-1} {\phi}}
  =  {\cal{G}}_{\rm c } \{ G_0^{-1} {\phi} \} - \frac{1}{2}
 ( {\phi} , G_0^{-1} {\phi}) 
 \; .
 \nonumber
 \\
 & &
 \label{eq:Gacdef}
 \end{eqnarray}
The amputated connected Green functions are then
 \begin{equation}
 G_{\rm ac}^{(n)} ( {\bf{r}}_1 , \ldots , {\bf{r}}_n )
  = \left. \frac{ \delta^{(n)} {\cal{G}}_{\rm ac} \{ {\phi} \}
}{ \delta {\phi} ( {\bf{r}}_1)  \ldots \delta {\phi} (
  {\bf{r}}_n )} \right|_{{\phi} = 0}
 \; . 
  \end{equation}
From the definition (\ref{eq:Gacdef})
and Eqs.(\ref{eq:Gcdef}) and (\ref{eq:GJdef}), it is easy to show that
${\cal{G}}_{\rm ac} \{ {\phi}\}$ has the following
functional integral representation,
 \begin{equation}
 e^{ {\cal{G}}_{\rm ac} \{ {\phi}\} } = 
 \frac{\int {\cal{D}} \{\phi^{\prime} \} e^{ - S_0 \{\phi^{\prime} \}
     -  S_{\rm int} \{ \phi^{\prime} +
   {\phi } \} }}{ 
 \int {\cal{D}} \{\phi^{\prime} \} e^{ - S_0 \{\phi^{\prime} \} } }
 \; .
 \end{equation} 
Furthermore, from the definition (\ref{eq:Gacdef}) and
the second line of Eq. (\ref{eq:Gcrepres})
we see that ${\cal{G}}_{\rm ac} \{ {\phi}\}$ has the following
representation in terms of a functional differential operator,
 \begin{equation}
 e^{ {\cal{G}}_{\rm ac} \{ \phi \}}
 =  e^{ \frac{1}{2} ( \frac{\delta}{\delta \phi} , G_0
   \frac{\delta}{\delta \phi }) }
  e^{ - S_{\rm int} \{ \phi \}}
\label{eq:Gacdiff}
 \; .
 \end{equation}

\subsection{Irreducible Green function}
\label{subsec:Girr}

The generating functional of the one-particle irreducible
Green functions contains the 
information about the many-body correlations
in the most compact form.
This generating functional 
is obtained from the generating functional of the connected Green
functions via a Legendre transformation,
 \begin{equation}
 {\cal{L}} \{ \varphi \} = ( \varphi , J) - {\cal{G}}_{\rm c} \{ J \{ \varphi
  \} \}
 \, ,
 \label{eq:Gammairdef}
 \end{equation}
where $J \{ \varphi \}$ is defined as a functional of the classical
field $\varphi$ via
 \begin{equation}
 \varphi ( {\bf{r}}) 
= \langle \phi ( {\bf{r}}) \rangle_{\rm c} = \frac{ \delta 
{\cal{G}}_{\rm c } \{ J \}}{\delta J ( {\bf{r}})
   } 
 \label{eq:phicldef}
 \; .
 \end{equation}
Hence, 
 \begin{eqnarray}
 {\cal{L}} \{ \varphi \} & = & F - F_0 + \sum_{n = 1}^{\infty}
 \frac{1}{n !} \int d {\bf{r}}_1 \ldots \int d {\bf{r}}_n
 L^{(n)} ({\bf{r}}_1 , \ldots , {\bf{r}}_n )
  \varphi  ( {\bf{r}}_1 ) \ldots \varphi ( {\bf{r}}_n )
 \; ,
 \nonumber
 \\
 & &
 \end{eqnarray}
where
 \begin{equation}
 L^{(n)} ( {\bf{r}}_1 , \ldots , {\bf{r}}_n )
  = \left. \frac{ \delta^{(n)} {\cal{L}} \{ \varphi \}
}{ \delta \varphi ( {\bf{r}}_1)  \ldots \delta \varphi (
  {\bf{r}}_n )} \right|_{\varphi = 0} 
 \; .
  \end{equation}
Note that in the absence of interactions ($S_{\rm int} = 0$)
the only non-zero vertex is
 \begin{equation}
 {L}_0^{(2)} ( {\bf{r}}_1 , {\bf{r}}_2 ) = G_0^{-1} ( {\bf{r}}_1 -
 {\bf{r}}_2) = \delta ( {\bf{r}}_1 - {\bf{r}}_2 )
 \left[ - \nabla_{ {\bf{r}}_2 }^2 + m_0^2 \right]
 \; .
\label{eq:G0rdef}
 \end{equation}
For the interacting system we have
by definition
 \begin{equation}
 {L}^{(2)} ( {\bf{r}}_1 , {\bf{r}}_2 ) = G_0^{-1} ( {\bf{r}}_1 -
 {\bf{r}}_2) + \Sigma ( {\bf{r}}_1 - {\bf{r}}_2 )
 \; ,
 \label{eq:gamma2}
 \end{equation}
where $\Sigma ( {\bf{r}})$ is the irreducible self-energy.
To obtain the generating functional of all  
irreducible $n$-point functions (including the two-point function),
we should therefore subtract the 
free propagator from Eq. (\ref{eq:Gammairdef}), defining
 \begin{eqnarray}
 {\cal{G}}_{\rm ir} \{ \varphi \} & = & {\cal{L}} \{ \varphi \} -
 \frac{1}{2} ( \varphi, G_0^{-1} \varphi ) =
( \varphi , J) -  \frac{1}{2} ( \varphi, G_0^{-1} \varphi )
- {\cal{G}}_{\rm c} \{ J \{ \varphi
  \} \}
 \, .
 \nonumber
 \\
 &  &
 \label{eq:Gammairdef2}
 \end{eqnarray}
The irreducible $n$-point functions are then
 \begin{equation}
 G_{\rm ir}^{(n)} ( {\bf{r}}_1 , \ldots , {\bf{r}}_n )
  = \left. \frac{ \delta^{(n)} {\cal{G}}_{\rm ir} \{ \varphi \}
}{ \delta \varphi ( {\bf{r}}_1)  \ldots \delta \varphi (
  {\bf{r}}_n )} \right|_{\varphi = 0} 
 \; .
  \end{equation}
%
%              R E F E R E N C E S

%
%
\end{document}